\documentclass[10pt,journal,compsoc,twoside]{IEEEtran}
\usepackage[utf8x]{inputenc}
\usepackage[american]{babel}
\usepackage{cite}
\usepackage{booktabs}
\usepackage{tabularx}
\usepackage{graphicx}
\usepackage{xspace}
\usepackage{makecell}
\usepackage{multirow}
\usepackage{adjustbox}
\usepackage[spacing,stretch=10,shrink=40]{microtype}
\usepackage[acronyms,nohypertypes={acronym},nomain]{glossaries}
\usepackage[binary-units, detect-all, per-mode=symbol, range-units=single, range-phrase=~to~]{siunitx}
\usepackage[dvipsnames]{xcolor}
\usepackage{amssymb}
\usepackage{amsbsy}
\usepackage{tikz}
\ifCLASSOPTIONcompsoc
  \usepackage[caption=false,font=normalsize,labelfont=sf,textfont=sf]{subfig}
\else
  \usepackage[caption=false,font=footnotesize]{subfig}
\fi

\usepackage[hidelinks]{hyperref}
\usepackage{cleveref}

\newif\ifthesis

\newif\ifpspin
\newif\ifllc
\newif\ifrevhl
\ifrevhl
  \colorlet{colorrevhl}{Blue}
  \newcommand{\revhl}[1]{\textcolor{colorrevhl}{#1}}
\else
  \colorlet{colorrevhl}{black}
  \newcommand{\revhl}[1]{#1}
\fi

\DeclareSIUnit{\GE}{GE}
\DeclareSIUnit{\dpflop}{dpflop}

\crefname{section}{\S}{\S\S}
\crefname{figure}{Fig.\@}{Figs.\@}
\crefname{table}{Table}{Tables}

\newacronym{ace}{ACE}{AXI Coherency Extensions}
\newacronym{asic}{ASIC}{application-specific integrated circuit}\glsunset{asic}
\newacronym{at}{AT}{area and timing}
\newacronym{axi}{AXI}{Advanced eXtensible Interface}
\newacronym{cdc}{CDC}{clock domain crossing}
\newacronym{chi}{CHI}{Coherent Hub Interface}
\newacronym{cpu}{CPU}{central processing unit}\glsunset{cpu}
\newacronym{cmo}{CMO}{cache management operation}
\newacronym{d2d}{D2D}{die-to-die link}
\newacronym{dma}{DMA}{direct memory access}
\newacronym{drc}{DRC}{design rule checking}\glsunset{drc}
\newacronym{dwc}{DWC}{data width converter}
\newacronym{eda}{EDA}{electronic design automation}
\newacronym{ff}{FF}{flip-flop}
\newacronym{fifo}{FIFO}{first-in first-out buffer}
\newacronym{flop}{FLOP}{floating-point operation}
\newacronym{fpga}{FPGA}{field-programmable gate array}\glsunset{fpga}
\newacronym{fpu}{FPU}{floating-point unit}
\newacronym{gf22}{GF22FDX}{GlobalFoundries' 22\,nm fully-depleted silicon-on-insulator}
\newacronym{gpgpu}{GPGPU}{general-purpose graphics processing unit}\glsunset{gpgpu}
\newacronym{io}{I/O}{input/output}
\newacronym[longplural={intellectual properties}]{ip}{IP}{intellectual property}
\newacronym{mlt}{MLT}{machine learning training}
\newacronym{mpsoc}{MPSoC}{multiprocessor system-on-chip}
\newacronym{msb}{MSB}{most significant bit}
\newacronym{nic}{NIC}{network interface controller}
\newacronym{nn}{NN}{neural network}
\newacronym[longplural={networks-on-chip}]{noc}{NoC}{network-on-chip}
\ifllc
  \newacronym{llc}{LLC}{last level cache}
\fi
\newacronym{lzc}{LZC}{leading-zero counter}
\newacronym{pcie}{PCIe}{Peripheral Component Interconnect Express}\glsunset{pcie}
\newacronym{qos}{QoS}{quality of service}
\newacronym{rtl}{RTL}{register-transfer level}
\newacronym[longplural={systems-on-chip}]{soc}{SoC}{system-on-chip}
\newacronym{soi}{SOI}{silicon-on-insulator}
\newacronym{spm}{SPM}{scratch-pad memory}
\newacronym{sram}{SRAM}{static random access memory}

\newcommand{\glsunknown}[1]{\glsreset{#1}\gls{#1}}
\newcommand{\glsunknownpl}[1]{\glsreset{#1}\glspl{#1}}
\newcommand{\glsknown}[1]{\glsunset{#1}\gls{#1}}
\newcommand{\glsknownpl}[1]{\glsunset{#1}\glspl{#1}}

\newcommand{\definition}[1]{\emph{#1}}
\newcommand{\valid}{\textit{valid}\xspace}
\newcommand{\ready}{\textit{ready}\xspace}
\newcommand{\flowrule}[1]{\textbf{(F#1)}}
\newcommand{\orderrule}[1]{\textbf{(O#1)}}
\newcommand{\bigO}[1]{\ensuremath{\mathcal{O}\left(#1\right)}}
\newcommand{\negcircnum}[1]{\includegraphics[height=.9em,trim=0 1.5 0 -1.5]{utf8_neg_circ_sans_digit_#1.pdf}}

\begin{document}

\bstctlcite{IEEEexample:BSTcontrol}

\setlength{\floatsep}{.4\baselineskip plus .2\baselineskip minus .2\baselineskip}
\setlength{\textfloatsep}{.5\baselineskip plus .2\baselineskip minus .3\baselineskip}
\setlength{\dblfloatsep}{\floatsep}
\setlength{\dbltextfloatsep}{\textfloatsep}
\setlength{\abovecaptionskip}{.125\baselineskip}

\newcommand{\titleraw}{An Open-Source Platform for High-Performance Non-Coherent On-Chip Communication}
\newcommand{\titlefmt}{\titleraw}
\title{\titlefmt}

\ifx\anonymous\undefined
\author{%
  Andreas~Kurth,~\IEEEmembership{Student Member,~IEEE,}
  Wolfgang~R\"{o}nninger, %
  Thomas~Benz, %
  Matheus~Cavalcante,~\IEEEmembership{Student Member,~IEEE,}
  Fabian~Schuiki, %
  Florian~Zaruba,~\IEEEmembership{Student Member,~IEEE,}
  and~Luca~Benini,~\IEEEmembership{Fellow,~IEEE}
}
\else
\author{\textit{Anonymous Author(s)}}
\fi
\markboth{%
}{%
    Kurth \MakeLowercase{\textit{et al.}}: \titleraw%
}

\IEEEtitleabstractindextext{%
\begin{abstract}
On-chip communication infrastructure is a central component of modern \glsunknownpl{soc}, and it continues to gain importance as the number of cores, the heterogeneity of components, and the on-chip and off-chip bandwidth continue to grow.
Decades of research on on-chip networks enabled cache-coherent shared-memory multiprocessors.
However, communication fabrics that meet the needs of heterogeneous many-cores and accelerator-rich \glspl{soc}, which are not, or only partially, coherent, are a much less mature research area.
In this work, we present a modular, topology-agnostic, high-performance on-chip communication platform.
The platform includes components to build and link subnetworks with customizable bandwidth and concurrency properties and adheres to a state-of-the-art, industry-standard protocol.
We discuss microarchitectural trade-offs and timing/area characteristics of our modules and show that they can be composed to build high-bandwidth (e.g., \SI{2.5}{\giga\hertz} and \SI{1024}{\bit} data width) end-to-end on-chip communication fabrics (not only network switches but also \acrshort{dma} engines and memory controllers) with high degrees of concurrency.
We design and implement
\ifpspin
\else
  a state-of-the-art ML training accelerator, where
\fi
our communication fabric scales to 1024 cores on a die, providing \SI{32}{\tera\byte\per\second} cross-sectional bandwidth at only \SI{24}{\nano\second} round-trip latency between any two cores.

\end{abstract}
}

\IEEEoverridecommandlockouts
\IEEEpubid{%
  \parbox{\textwidth}{%
    \centering
    10.1109/TC.2021.3107726~\copyright~2021~IEEE.
    Personal use is permitted, but republication/redistribution requires IEEE permission.\\
    See \url{https://www.ieee.org/publications/rights/index.html} for more information.
  }%
}
\maketitle
\IEEEpubidadjcol

\IEEEdisplaynontitleabstractindextext

\ifthesis\else
  \IEEEraisesectionheading{\section{Introduction}\label{sec:introduction}}
\fi

\IEEEPARstart{O}{n-chip networks} are the primary means of communication inside modern multi- and many-core processing \glspl{soc}~\cite{jerger2017,dally2003nocs,benini2006nocs,kundu2018nocs}.
As the number of cores, the heterogeneity of components, and the on- and off-chip bandwidth continue to grow to meet ever higher application demands, on-chip networks continue to gain importance.
Decades of research on on-chip networks were instrumental for %
breakthroughs in scalability of homogeneous shared-memory multiprocessors,
and a continuation of this research is necessary to realize the full potential of many-core accelerators and accelerator-rich heterogeneous \glspl{soc}.

\newcommand{\goal}[1]{\textbf{(G#1)}}
Ideally, \gls{soc} designers could compose on-chip networks from a platform of components according to the requirements of their application.
The central design goals of such a platform are:
\ifthesis\begin{itemize}\fi
\newcommand{\goalitem}[1]{%
  \ifthesis%
    \item[\goal{#1}]%
  \else%
    \goal{#1}%
  \fi%
}
\goalitem{1} Elementary, modular components that can implement any topology and that separate concerns such as routing and buffering.
\goalitem{2} Parametrizable components (e.g., data width, transaction concurrency) to cover a large design space.
\goalitem{3} Bridging components to connect heterogeneous \gls{soc} elements (e.g., GPU SMs, \acrshort{dma} engines, and domain-specific accelerators) and their subnetworks, each with unique, application-driven latency and bandwidth requirements.
\goalitem{4} Compliance with an industry-standard protocol for extensibility, third-party compatibility, and verifiability.
\goalitem{5} Detailed characterization of the complexity and trade-offs of the components in terms of performance vs.\ cost (area, power) %
to guide design and optimization efforts.
\ifthesis\end{itemize}\fi

Commercial offerings that meet (parts of) these goals exist from multiple vendors (details in \cref{sec:related_work}), but their microarchitecture, complexity, and performance are well-guarded trade secrets.
Research has also worked toward those goals (details in \cref{sec:related_work}), but, to the best of our knowledge,
an end-to-end platform for non-coherent on-chip communication that meets the needs of heterogeneous \glspl{soc} has not been presented yet in open literature and is not available as open-source hardware.

In this work, we fill this gap with these contributions:
\begin{enumerate}
    \item
        We present a modular, topology-agnostic~\goal{1}, high-performance on-chip communication platform of parametrizable components~\goal{2} for a state-of-the-art, industry-standard %
        protocol~\goal{4}~(\cref{sec:arch}).
        The components include bridges and converters to link subnetworks with different bandwidth and concurrency properties~\goal{3}.
        We publish the modules of our platform, implemented in industry-standard SystemVerilog, under a permissive open-source license for research and industrial usage.
    \item
        We discuss microarchitectural trade-offs and timing/area characteristics of the modules in our platform~\goal{5}, both theoretically/asymptotically and with topographical synthesis results~(\cref{sec:impl}).
        We show that our modules can be composed to build high-bandwidth (e.g., \SI{2.5}{\giga\hertz} and \SI{1024}{\bit} data width), end-to-end on-chip communication fabrics (e.g., \acrshort{dma} engine to memory controller), with high degrees of concurrency (e.g., up to 256 independent concurrent transactions) and flexibility (e.g., 64-bit subnetworks).%
    \item
        We design and implement (post-P\&R) %
        a state-of-the-art many-core \gls{mlt} accelerator in a modern \SI{22}{\nano\meter} technology~(\cref{sec:sys}), where
        our communication fabric scales to 1024 cores on a die, which deliver more than \SI{2}{\tera\dpflop\per\second}, providing \SI{32}{\tera\byte\per\second} cross-sectional bandwidth at only \SI{24}{\nano\second} round-trip latency between any two cores.
\end{enumerate}
We focus on non-coherent on-chip communication for two main reasons:
First, coherent on-chip communication in homogeneous many-core processors has been studied extensively (see \cref{sec:related_work} for an overview).
Second, many complex heterogeneous \glspl{soc} (e.g., mobile application \glspl{soc}~\cite{snapdragon865}, high-speed networking \glspl{soc}~\cite{tomahawk4}) and massively parallel data processing architectures (e.g., \glspl{gpgpu}~\cite{smith2020ampere}) are not or only partially cache-coherent.

This paper is organized as follows:
We present the architecture of our on-chip communication platform in \cref{sec:arch} and characterize its performance and complexity in \cref{sec:impl}.
We then use our platform to design, implement, and evaluate the communication fabric of %
a state-of-the-art many-core \gls{mlt} accelerator in \cref{sec:sys}.
Finally, we compare with related work in \cref{sec:related_work} and conclude in \cref{sec:conclusion}.

\section{Architecture}%
\label{sec:arch}

Current on-chip communication is centered around the premise of high-bandwidth point-to-point data transfers.
To fulfill this premise despite increasing point-to-point latency, three central traits of current on-chip communication protocols are: \emph{burst-based transactions}, \emph{multiple outstanding transactions}, and \emph{transaction reordering}.
Our design targets these central traits in general, so the concepts we present potentially apply to a wide range of modern on-chip protocols.
More tangibly, we adhere to the latest revision (5) of the AMBA \gls{axi}~\cite{axi}.
\Gls{axi} is one of the industry-dominant protocols and the only protocol with an open, royalty-free specification and a widespread adoption in current systems designed by many different companies.
Other protocols with similar properties are discussed in \cref{sec:related_work}.

\revhl{
\subsubsection*{Terminology and Protocol Essentials}
A \definition{module} is a distinct functional unit that has at least one on-chip network port.
A \definition{port} is a collection of input and output signals of a module.
A port can be either a \definition{master port}, on which the module initiates transactions, or a \definition{slave port}, on which the module responds to transactions.
One module can have multiple slave and master ports.
We collectively call the five independently-handshaked channels connecting a master port to a slave port a \definition{bundle}.
Each \definition{channel} consists of multiple isodirectional payload signals and two signals for bi-directional flow control.
A \definition{beat} is the data transferred on one channel upon one handshake; it is the smallest unit of communication.
}
We focus on \valid-\ready flow control, where the channel master drives the \valid signal and the payload signals and the channel slave drives the \ready signal (but other flow control schemes, e.g., credit-based, are possible).
A \definition{handshake} occurs when \valid and \ready are high on a rising clock edge.
\ifthesis%

\fi%
There are two essential rules in \valid-\ready flow control:
\ifthesis\begin{itemize}\fi%
\newcommand{\flowruleitem}[1]{%
  \ifthesis%
    \item[\flowrule{#1}]%
  \else%
    \flowrule{#1}%
  \fi
}
\flowruleitem{1}~Stability Rule: Once \valid is high, \revhl{\valid} and the payload must not change until \revhl{the handshake occurs}.
\flowruleitem{2}~Acyclicity Rule: The channel slave may depend on \valid to be high before setting \ready high, but the channel master may not depend on \ready to be high before setting \valid high.
\ifthesis\end{itemize}\fi%

\begin{figure}
    \centering
    \includegraphics[width=\columnwidth]{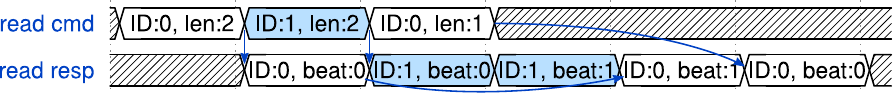}
    \caption{
        Transaction ID and ordering example.
        Three read commands are issued; the first and last have the same ID, the one between has a different ID.
        This situation might occur, e.g., if the transactions are issued by different masters.
        The first response beat is for the first command.
        After that follow both response beats for the second command.
        This interleaving of response beats is allowed because the first two commands have different IDs, so their responses may be interleaved.
        Then follows the second response beat for the first command.
        The response beat for the last command comes at the end, because it must not come before the last response beat of the first command (which has the same ID).
    }%
    \label{fig:arch:id_example}
\end{figure}

Each transaction has a \definition{direction} (read or write):
A \definition{write transaction} starts with one beat on the \definition{write command channel} followed by one or multiple beats on the \definition{write data channel} and ends with a single beat on the \definition{write response channel}.
A \definition{read transaction} starts with one beat on the \definition{read command channel} and ends with one or the last of multiple beats on the \definition{read response channel}.
\revhl{A transaction is \definition{outstanding} in the time interval starting with the handshake of the command beat and ending with the handshake of the (last) response beat.}
Each transaction has a numeric \definition{ID}. %
\ifthesis%

\fi%
IDs define the order of transactions and beats according to the following rules:
\ifthesis\begin{itemize}\fi%
\newcommand{\orderruleitem}[1]{%
  \ifthesis%
    \item[\orderrule{#1}]%
  \else%
    \orderrule{#1}%
  \fi
}
\orderruleitem{1}~Inter-Transaction Ordering: Any two transactions in the same direction and ID are ordered.
\orderruleitem{2}~Response Ordering: Any two responses with the same direction and ID must be in the same order as their commands.
\orderruleitem{3}~Write Beat Ordering: Write data beats do not have an ID and are therefore always ordered.
\ifthesis\end{itemize}\fi%
An example of IDs and their ordering is shown in \cref{fig:arch:id_example}.
\ifthesis
\begin{table}
    \centering
    \begin{tabular}{ l l l }
        \toprule
        \thead{Category} & \thead{Module} & \thead{Section} \\
        \midrule
        \multirowcell{2}[0pt][l]{Elementary\\Components} & Network Multiplexer & \ref{sec:arch:mux} \\
        {} & Network Demultiplexer & \ref{sec:arch:demux} \\
        \midrule
        \multirowcell{2}[0pt][l]{Network\\Junctions} & Crossbar & \ref{sec:arch:xbar} \\
        {} & Crosspoint & \ref{sec:arch:xp} \\
        \midrule
        \multirowcell{2}[0pt][l]{Concurrency\\Control} & ID Remapper & \ref{sec:arch:id_remapper} \\
        {} & ID Serializer & \ref{sec:arch:id_serializer} \\
        \midrule
        \multirowcell{2}[0pt][l]{Data Width\\Converters} & Data Upsizer & \ref{sec:arch:dw_converter:upsizer} \\
        {} & Data Downsizer & \ref{sec:arch:dw_converter:downsizer} \\
        \midrule
        Clock Domain Crossing & Clock Domain Crossing & \ref{sec:arch:cdc} \\
        \midrule
        Data Movement & \acrshort{dma} Engine & \ref{sec:arch:dma_engine} \\
        \midrule
        \ifllc%
          \multirowcell{3}[0pt][l]{On-Chip Memory\\Endpoints}
        \else%
          \multirowcell{2}[0pt][l]{On-Chip Memory\\Endpoints}
        \fi%
        & Simplex Memory Controller & \ref{sec:arch:on-chip_mem_ctrl:simplex} \\
        & Duplex Memory Controller & \ref{sec:arch:on-chip_mem_ctrl:duplex} \\
        \ifllc%
          & Last Level Cache & \ref{sec:arch:llc} \\
        \fi%
        \bottomrule
    \end{tabular}
    \caption{Overview of the modules in our on-chip communication platform.}%
    \label{tbl:arch:overview}
\end{table}
\fi
\ifthesis%
\par
\fi%
\ifthesis%
An overview of the modules in our on-chip communication platform is given in \cref{tbl:arch:overview}.
\fi%
\ifthesis%
In this section, we discuss their
\else%
The rest of this section discusses the
\fi%
microarchitecture and design trade-offs
\ifthesis\else%
of our on-chip communication platform%
\fi%
, from elementary components through all essential interconnecting modules to endpoints of increasing complexity.

\ifthesis%
  \subsection{Elementary\texorpdfstring{\,}{ }Components: Network\texorpdfstring{\,}{ }(De)Muxes}%
\else%
  \subsection{Elementary\texorpdfstring{\,}{ }Components: Network\texorpdfstring{\,}{ }(De)Multiplexers}%
\fi%
\label{sec:arch:mux_demux}

Our network multiplexers and demultiplexers are the elementary components that join multiple ports to one and split one port into multiple, respectively. %
In doing so, they must adhere to the relations between the channels and to the ordering rules \orderrule{1--3}.
They are obviously used to build network junctions (e.g., crossbars), but they can be reused far beyond that because they implement a central part of the communication protocol.
In fact, these elementary components are essential for almost all modules of our platform \revhl{and can be used to design custom endpoints without having to deal with all protocol intricacies.}
\revhl{Each of our \emph{network} (de)multiplexers contains simple \emph{logic} (de)multiplexers, but it also contains other components to implement the protocol.
In the remainder of this article, `(de)multiplexer' refers to a \emph{network} (de)multiplexer unless explicitly preceded by `logic'.
}

\subsubsection{Network Multiplexer}%
\label{sec:arch:mux}

The multiplexer, which connects multiple slave ports to one master port, consists of multiplexing components for the forward channels and demultiplexing components for the backward channels.
The complexity lies in \emph{de}multiplexing the backward channels, because the multiplexer needs the information to which output a beat on a backward channel must be routed.
Multiplexing the command channels simply requires the selection of a \valid beat, with the restriction that a selection must be stable once made \flowrule{1}.

\begin{figure}
    \centering%
    \includegraphics[width=\columnwidth]{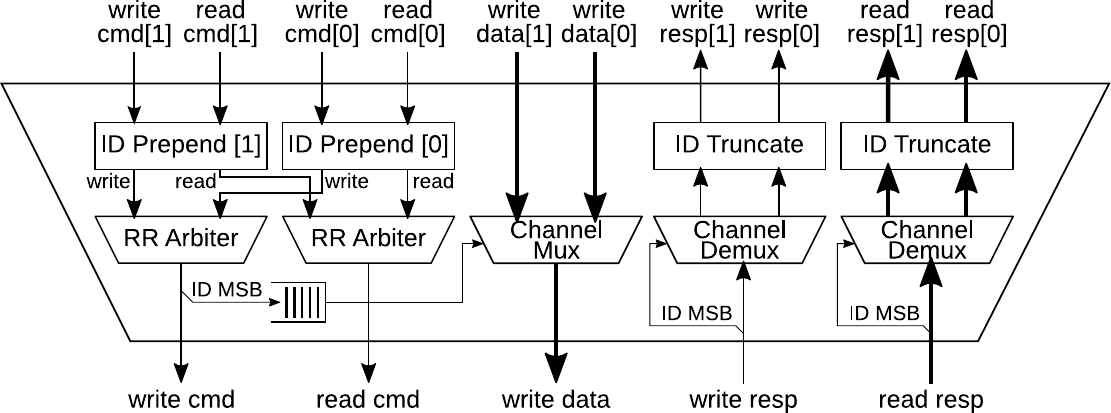}%
    \caption{%
      Architecture of our network \textbf{multiplexer}, drawn with two slave ports (at the top).
      \revhl{%
        Here and in \cref{fig:arch:demux}, `Channel (De)mux' refer to logic (de)multiplexers with additional handshake logic for one channel.
        In all later figures, `Mux' refers to the entire module in \cref{fig:arch:mux} and `Demux' refers to the entire module in \cref{fig:arch:demux}. 
    }
    }%
    \label{fig:arch:mux}%
\end{figure}

Our multiplexer architecture is shown in \cref{fig:arch:mux}.
We first prepend the ID of each command beat with the number of the slave port.
We then select among beats on the command channels with round-robin (RR) arbitration trees.
For writes, the decision is forwarded through a \gls{fifo} to a multiplexer for the write data beats, which is sufficient due to \orderrule{3}.
As commands out of our multiplexer carry the input port information in the \glspl{msb} of their ID, routing responses is as simple as demultiplexing based on the \glspl{msb} and then truncating the ID to the original width.
Another key advantage is that transactions with the same ID from any two different slave ports remain independent, so \orderrule{1} does not restrict communication through our multiplexer.
Note that \emph{channel demultiplexing} means the payload is the same for all demux outputs and only the handshake signals are (de)multiplexed.

Alternative multiplexer architectures could do without extending the ID, for example by allowing only transactions with different IDs concurrently or by remapping IDs internally.
However, the former restricts communication, and the latter significantly increases the complexity of the multiplexer.
Nonetheless, some network modules grow exponentially in complexity with the ID width.
We have a modular solution to this challenge with the ID width converters discussed in~\cref{sec:arch:iw_converter}.

\subsubsection{Network Demultiplexer}%
\label{sec:arch:demux}

The demultiplexer, which connects one slave port to multiple master ports, is more complex than the multiplexer due to the ordering rules:
When the demultiplexer gets two commands with the same ID and direction \orderrule{1} that go to two different master ports, it must deliver the corresponding responses in the same order \orderrule{2}.
After the demultiplexer, however, transactions on different master ports are independent, so the demultiplexer cannot rely on the order of downstream responses to fulfill \orderrule{2}.

\begin{figure}
    \centering%
    \includegraphics[width=\columnwidth]{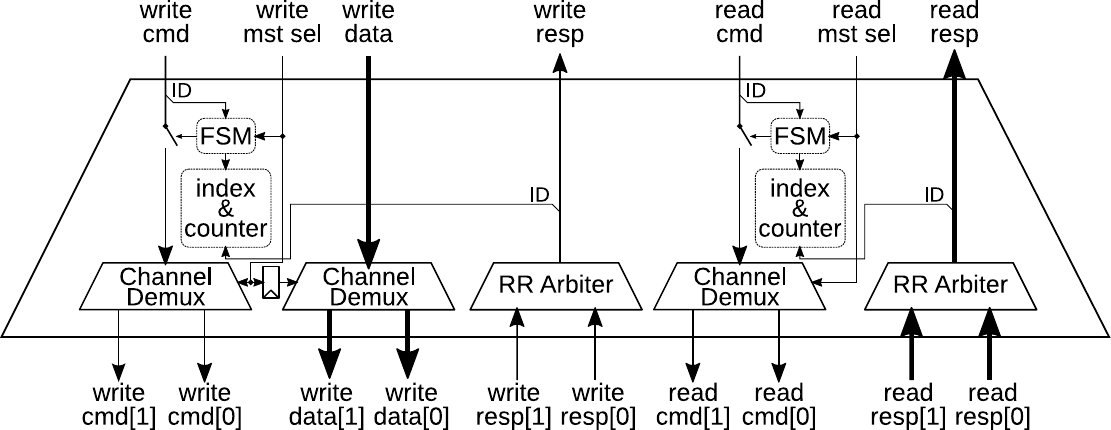}%
    \caption{%
      Architecture of our network \textbf{demultiplexer}, drawn with two master ports (at the bottom).
    }%
    \label{fig:arch:demux}%
\end{figure}

Our demultiplexer architecture, shown in \cref{fig:arch:demux}, solves this by enforcing that all concurrent transactions with the same direction and ID target the same master port.
For example, when a write with ID $A$ targets master port $0$, it is only forwarded if no writes with ID $A$ to master ports other than $0$ have outstanding responses; otherwise, the write must wait.
To track this information, the demultiplexer contains one counter and one index register per ID and direction.
Commands that fulfill the aforementioned requirement increase the counter; the (last) response decreases the counter.
A channel register between the write command channel and the demultiplexer of the write data channel stores the master port index of an ongoing write burst while the command channel is independently handshaked \flowrule{1}.
Write commands and data bursts are sent in lockstep due to \orderrule{3}; without this restriction, the write command and data channels could deadlock downstream. %
The multiple read and write response channels are joined through a round-robin arbitration tree.

Alternative demultiplexer architectures could do without requiring all concurrent transactions with the same direction and ID to target the same master port, for example by remapping IDs internally.
However, this significantly increases the complexity of the demultiplexer, which would have to reorder responses internally to fulfill \orderrule{2}.
Instead of introducing this complexity, we let a master use different IDs for different endpoints if it can handle out-of-order responses.

\revhl{%
Compared with a 1-to-N crossbar, the demultiplexer has a fundamental advantage concerning how transactions are routed:
With the crossbar, the \emph{address} of a transaction determines to which master port it is routed.
With the demultiplexer, the \emph{select inputs} (one for reads, one for writes) determine to which master port a transaction is routed.
This means a module instantiating the demultiplexer can freely decide which submodule handles a transaction.
That decision does not even have to be based on the properties of the transaction but could, for example, be a function of the state of the module.
This difference implies that the demultiplexer is a more universal elementary component than a 1-to-N crossbar.
}

\revhl{%
\emph{Logic} demultiplexers are so universally used in digital circuits that our \emph{network} demultiplexer may seem like a trivial sequel.
However, as the architecture depicted in \cref{fig:arch:demux} and described in this section shows, our demultiplexer handles crucial and complex parts of the protocol (\orderrule{1--3}, \flowrule{1}).
Thus, even though our demultiplexer simply takes a select signal to route transactions, it unburdens user modules from dealing with intricacies of the protocol while it enables them to arbitrarily route transactions to submodules or ports.
}

\subsection{Network Junctions: Crossbars and Crosspoints}%
\label{sec:arch:junction}

\subsubsection{Crossbar}%
\label{sec:arch:xbar}

\begin{figure}
    \centering%
    \newlength{\axixbarwidth}%
    \ifthesis%
      \setlength{\axixbarwidth}{.95\columnwidth}%
    \else%
      \setlength{\axixbarwidth}{.875\columnwidth}%
    \fi%
    \includegraphics[width=\axixbarwidth]{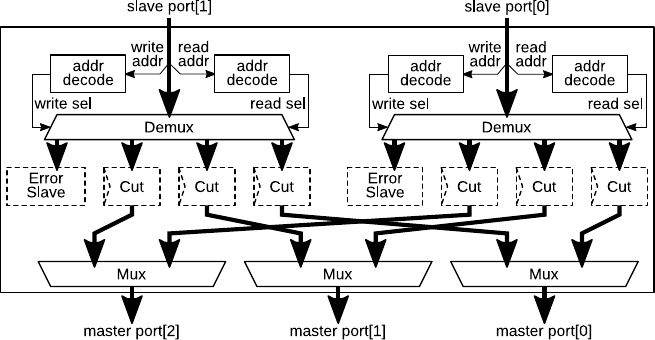}%
    \caption{%
      Architecture of our \textbf{crossbar}, drawn with two slave and three master ports.
      Each fat arrow represents a \revhl{bundle, with the arrow head pointing in the direction of the command channels}.
      Components with dashed outline are optional.
    }%
    \label{fig:arch:xbar}%
\end{figure}

The elementary components in \cref{sec:arch:mux_demux} can be combined to form a fully-connected crossbar, shown in \cref{fig:arch:xbar}, where each slave port has a dedicated connection to each master port.%

At each slave port, two address decoders (one for reads, one for writes) drive the selection signals of a demultiplexer.
In the standard configuration, all slave ports use the same addresses for one master port, but different configurations would be possible.
There are two alternatives for handling transactions to an address that is not defined in a decoder.
First, one master port can be defined as default port.
This is useful, for example, in a hierarchical topology where each downlink has a specific range of addresses and any address outside the downlink addresses is sent to higher hierarchy levels through the uplink.
Second, one can instantiate an error slave, which terminates all transactions with protocol-compliant error responses.
These two alternatives can be selected per slave port with a synthesis parameter.

Optional pipeline registers can be inserted on all or some of the five channels of each internal bundle.
These registers cut all combinational signals (including handshake signals), thereby  adding a cycle of latency per channel and pipelining the crossbar so its critical path is no longer than that of the demultiplexer or multiplexer.
These pipeline registers can be added without risking deadlocks, but this is not trivial:
Of the four Coffman conditions~\cite{coffman1971}, (1) Mutual Exclusion is fulfilled on the write data channel after the multiplexer, (2) Hold and Wait is fulfilled as each pipeline register must hold its value once filled, (3) No Preemption is fulfilled by \orderrule{3} on the write data channel, and (4) Circular Wait would be fulfilled by round-robin arbitration of write command and data beats.
However, the demultiplexer breaks condition (4) by restricting write commands to be issued in lockstep with write data bursts (i.e., the next write command is only issued after the previous write data burst has completed), thereby preventing deadlocks despite pipeline registers, which introduce condition (2).

\subsubsection{Crosspoint}%
\label{sec:arch:xp}

As the multiplexers in the crossbar expand the ID width, the master ports of the crossbar have a wider ID than the slave ports.
This prevents the direct use of our crossbar as nodes in a regular on-chip network where each node (also called ``router'' \revhl{or ``switch''}) has isomorphous slave and master ports.
To solve this problem, we introduce a crosspoint.

\begin{figure}
    \centering
    \newlength{\axixpwidth}%
    \ifthesis%
      \setlength{\axixpwidth}{.7\columnwidth}%
    \else%
      \setlength{\axixpwidth}{.5\columnwidth}%
    \fi%
    \includegraphics[width=\axixpwidth]{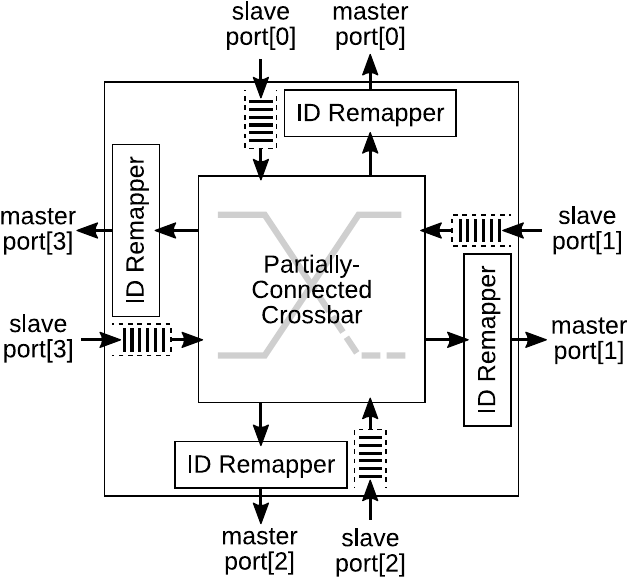}
    \caption{%
      Architecture of our \textbf{crosspoint}, drawn with four slave and master ports.
      Each arrow represents a \revhl{bundle, with the arrow head pointing in the direction of the command channels}.
      The input queues are optional.
    }
    \label{fig:arch:xp}
\end{figure}

Our crosspoint, shown in \cref{fig:arch:xp}, has three additional properties over the crossbar that make it better suited for composing arbitary regular on-chip topologies.
First, it contains a crossbar that is not necessarily fully connected:
The connection between any slave and master port can be omitted with a synthesis parameter.
This is useful to prevent routing loops when a module has both a master and a slave port into the crosspoint, and it minimizes the physical resources on links that would be unused.
All flow and arbitration control logic of the crosspoint is inside the crossbar.
Second, the crosspoint contains an ID remapper (\cref{sec:arch:id_remapper}) on each master port, which reduces the ID width to that of the slave ports.
Thus, the slave and master ports of each crosspoint are isomorphous.
Third, an input queue of configurable depth can be enabled for each slave port to reduce backpressure in mesh topologies.

\subsection{Concurrent Transactions: ID Width Converters}%
\label{sec:arch:iw_converter}

The ID of transactions is central to their ordering \orderrule{1--2}.
Essentially, the commands and responses of any two transactions can be independently reordered if they have different IDs.
This makes a high number of possible IDs attractive to prevent bottlenecks due to ordering constraints.
However, tracking a high number of IDs is complex for network components (e.g., demultiplexer \cref{sec:arch:demux,sec:impl:demux}).

ID width converters are the on-chip network designer's instrument to balance the number of independent concurrent transactions vs.\ circuit complexity.
We focus on \emph{reducing} the ID width (as extending it is trivial).
There are two first-order parameters for ID reduction: the width of IDs at the output, $O$, and the maximum number of unique IDs at the input, $U$.
The relation between $O$ and $U$ determines whether all transactions that were independent at the input remain independent at the output:
If $U \leq 2\sp{O}$, every unique ID at the input can be represented by a unique ID at the output, therefore retaining transaction independence.
This means the sparsely used input ID space can be `compressed' to a narrower, densely used output ID space by \emph{remapping} IDs (\cref{sec:arch:id_remapper}).
If $U > 2\sp{O}$, there are not enough output IDs to represent all $U$ unique IDs.
This means some transactions with originally different IDs will have to be mapped to the same ID, thereby \emph{serializing} them (\cref{sec:arch:id_serializer}).

\subsubsection{ID Remapper}%
\label{sec:arch:id_remapper}

\begin{figure}
    \centering%
    \includegraphics[width=\columnwidth]{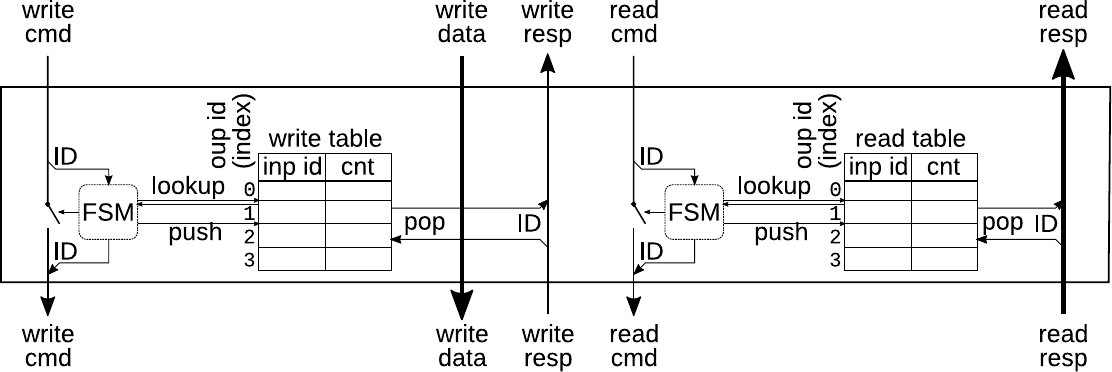}%
    \caption{%
      Architecture of our \textbf{ID remapper}, drawn with up to four unique concurrent IDs (per direction).
    }%
    \label{fig:arch:id_remapper}%
\end{figure}

Our ID remapper, shown in \cref{fig:arch:id_remapper}, remaps IDs with one table per direction.
The table has as many entries as there are unique input IDs, and it is indexed by the output ID.
Each table entry has two fields: the input ID and a counter that records how many transactions with the same ID are in flight.
The counter is incremented on command handshakes and decremented on (last) response handshakes.
The mapping from input to output IDs is injective.
Obtaining the input ID from an output ID (to remap responses) is as simple as indexing the table.
Determining the output ID for an input ID (to remap commands) requires a comparison of the input ID to all IDs in the table.
If the table currently contains an entry for the input ID, the same output ID must be used \orderrule{1}.
If the table does not currently contain an entry for the input ID, the output ID is the index of the next free table entry.

Alternative ID remapper architectures could feature an additional table indexed by input IDs to look output IDs up.
However, under the assumption of the remapper that the input ID space is sparse, such an additional table would be mostly empty.
Therefore, it would be a poor usage of hardware resources and we omit it at the cost of a longer ID translation path, which could be pipelined.

\subsubsection{ID Serializer}%
\label{sec:arch:id_serializer}

\begin{figure}
    \centering%
    \includegraphics[width=\columnwidth]{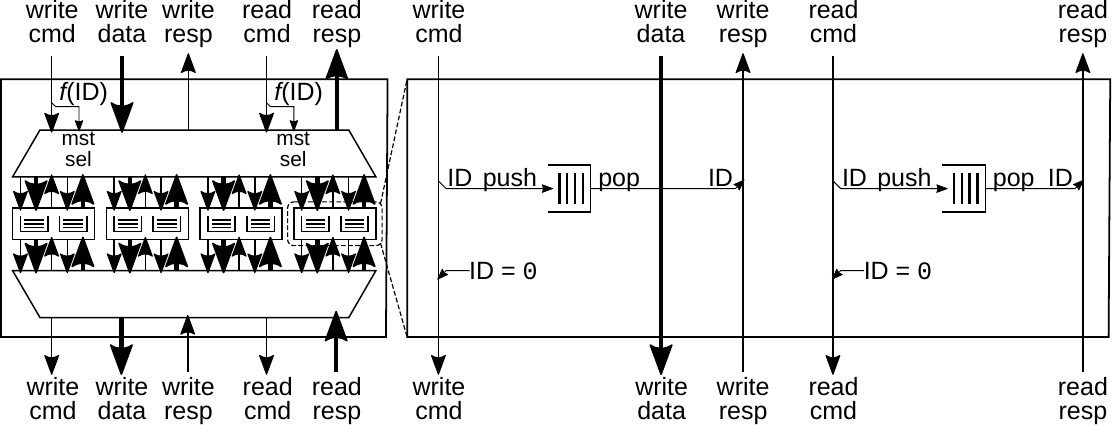}%
    \caption{%
      Architecture of our \textbf{ID serializer}, drawn with four master port IDs (per direction).
    }%
    \label{fig:arch:id_serializer}%
\end{figure}

If the number of unique IDs at the input of the ID width converter, $U$, exceeds the number of available IDs at the output, $2\sp{O}$, both the input and the output ID space are densely used.
In this case, it is not possible to retain the uniqueness of all IDs during conversion, and we call the transformation that imposes additional ordering \emph{serialization}.
Serialized transactions still have concurrently outstanding commands, but they are now required to be handled in-order.

Our ID serializer, shown in \cref{fig:arch:id_serializer}, transforms IDs with one \gls{fifo} per direction and master port ID.
At the slave port of the serializer, a demultiplexer assigns commands to one of the \gls{fifo} submodules through a combinational function $f$ of the transaction ID (e.g., the ID modulo the number of master port IDs).
The demultiplexer is a reduced configuration of our network demultiplexer (\cref{sec:arch:demux}) without ID counters because $f$ assigns identical IDs to the same master port (and thus the same output ID \orderrule{1}).
In each \gls{fifo} submodule, the ID of a command is pushed into a \gls{fifo} and then truncated to zero.
This \gls{fifo} reflects the transaction ID in responses \orderrule{2}, and the last response of a transaction pops from the \gls{fifo}.
After the \glspl{fifo}, an instance of our network multiplexer (\cref{sec:arch:mux}) assigns each transaction the index of its \gls{fifo} and merges the commands to the single master port of the ID serializer.

Alternative ID serializer architectures could use one memory where one linked list per master port ID is stored for ID reflection.
This would allow to dynamically grow queues in memory rather than statically provisioning hardware resources to accommodate a fixed maximum of transactions per master port ID. %
However, pushing and popping IDs from this memory is on the critical path of the serializer, so we prefer the architecture with multiple \glspl{fifo}.

\subsection{Data Width Converters}%
\label{sec:arch:dw_converter}

The data width of network components depends on their bandwidth requirements.
For instance, the master port of a high-performance \acrshort{dma} engine might have \SI{512}{\bit} data width while that of a 64-bit processor core typically has \SI{64}{\bit}.
This extends to subnetworks, e.g., separate networks for the \acrshort{dma} engine and the cores.
However, as subnetworks with different data widths are joined, e.g., at endpoints such as memories, 
\glspl{dwc} are required to convert between data widths.
\glspl{dwc} can be either \emph{upsizers}, converting from narrow to wide, %
or \emph{downsizers}, converting from wide to narrow. %
Although similar in purpose, up- and downsizer are not fully symmetric.
In fact, the upsizer has higher performance requirements than the downsizer, since it must utilize the higher-bandwidth network as much as possible to minimize the impact on other components on the high-bandwidth network. %

\subsubsection{Data Upsizer}%
\label{sec:arch:dw_converter:upsizer}

\begin{figure*}
    \centering
    \includegraphics[width=\textwidth]{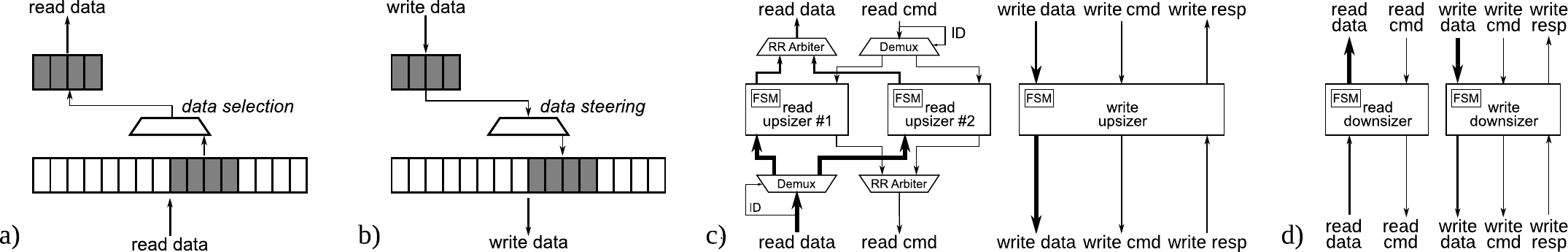}
    \caption{%
        Architecture of our \textbf{\acrfullpl{dwc}}.
        (a) Data selection in the read response and (b) data steering in the write data channel of the upsizer.
        (c) Upsizer, drawn with two outstanding read transactions.
        (d) Downsizer.
    }
    \label{fig:arch:dwc}
\end{figure*}

A data upsizer has a narrow slave port of data width $D\sb{\text{N}}$ and a wider master port of data width $D\sb{\text{W}}$.
In the simplest operating mode, pass-through, the upsizer %
does lane selection on read responses (\cref{fig:arch:dwc}a), selecting a slice of a wide incoming word, and lane steering on write data, aligning narrow incoming data into the wider outgoing word (\cref{fig:arch:dwc}b).
In pass-through mode, the upsizer does not change the number of bytes transferred in each beat.
This can be required by transaction attributes (e.g., to device memory).
In terms of performance, however, this underutilizes the high-bandwidth network, which inherits the throughput of its low-bandwidth counterpart.
Utilization can be increased by reshaping incoming bursts with many narrow beats into bursts with fewer wide beats:
several narrow write data beats are packed into one wide beat, and one wide read response beat is serialized into several narrow beats.

Our data upsizer, shown in \cref{fig:arch:dwc}c, is capable of upsizing between interfaces of any data width.
It is composed by two modules, read and write upsizers, that perform lane selection and steering, besides deciding whether to upsize the transaction based on its properties.
Due to~\orderrule{3}, only one write upsizer is needed, containing a buffer of width $D\sb{\text{W}}$ to perform data packing.
On the read response channel, the data upsizer handles a certain number of outstanding read transactions in parallel.
Each incoming read transaction is assigned an idle read upsizer, unless there is an active upsizer handling a transaction with the same ID.
For that case, we ensure~\orderrule{1} by enforcing that incoming transactions with the same ID are handled by the same read upsizer.
Each read upsizer has a $D\sb{\text{W}}$ buffer to hold incoming beats.
This avoids blocking the wide read response channel during serialization.

\subsubsection{Data Downsizer}%
\label{sec:arch:dw_converter:downsizer}

A data downsizer has a wide slave port of data width $D\sb{\text{W}}$ and a narrower master port of data width $D\sb{\text{N}}$.
In the simplest operating mode, pass-through, the downsizer does steering on the read data channel and selection on the write data channel, symmetrical to the base operations of the data upsizer.
Our downsizer, shown in \cref{fig:arch:dwc}d, differs from the upsizer in two key points:
First, the downsizer has lower performance requirements than the data upsizer, since %
it connects to a lower-bandwidth subnetwork, e.g., peripherals.
This means it does not need to support multiple outstanding reads.
Second, when downsizing, the downsizer converts few wide beats into multiple narrow beats.
It is possible that the resulting burst is longer than the longest buffer allowed by the protocol.
In this case, the downsizer needs to break the incoming burst into a sequence of bursts.
To handle this corner case, among others, the control logic of the read and write downsizers is more complex than those in the upsizer.

\color{colorrevhl}%
\subsection{Clock Domain Crossing}%
\label{sec:arch:cdc}

\begin{figure}
    \centering%
    \includegraphics[width=\columnwidth]{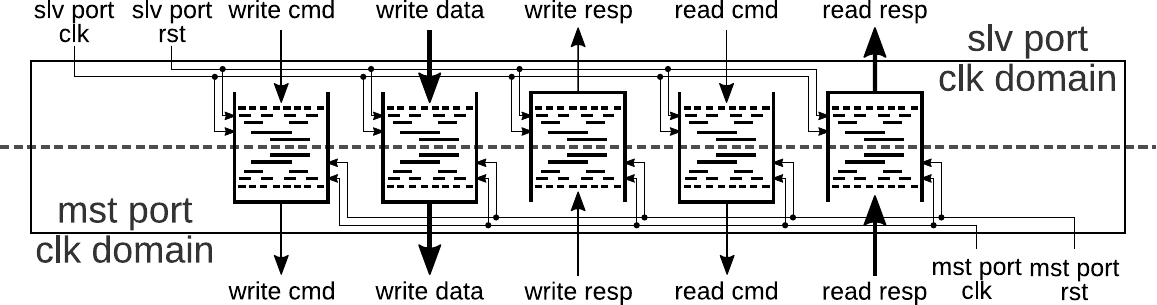}%
    \caption{%
      \revhl{%
      Architecture of our \textbf{\acrfull{cdc}}.
      Each channel goes through a \acrshort{cdc} \acrshort{fifo}, which has two Gray-coded counters.%
      }
    }%
    \label{fig:arch:cdc}%
\end{figure}

A on-chip network can span multiple clock domains, yet all our modules have a single clock input\footnote{\revhl{%
  For modules with a single clock input, we do not draw the clock input and clock wires to all sequential cells in the block diagram in order to not overcrowd the diagram.
}} -- except one: the \gls{cdc} has two clock inputs, one to which all signals of its slave port are synchronous and one for its master port.
The \gls{cdc} can be placed between any two modules in different clock domains.
This enables the creation of independently-clocked subnetworks as well as the connection to endpoints that provide their own clock.
In our \gls{cdc}, shown in \cref{fig:arch:cdc}, each channel goes through a \gls{cdc} \gls{fifo}, which has two Gray-coded counters: one for pushing the \gls{fifo} in one clock domain and one for popping from the \gls{fifo} in the other clock domain.
The implementation follows well-established \gls{cdc} principles~\cite{apperson2007,cummings2008,strano2010}.%
\color{black}%

\vspace{0ex plus 1.5ex}
\subsection{Data Movement: DMA Engine}%
\label{sec:arch:dma_engine}

\begin{figure*}
    \centering
    \includegraphics[width=\textwidth]{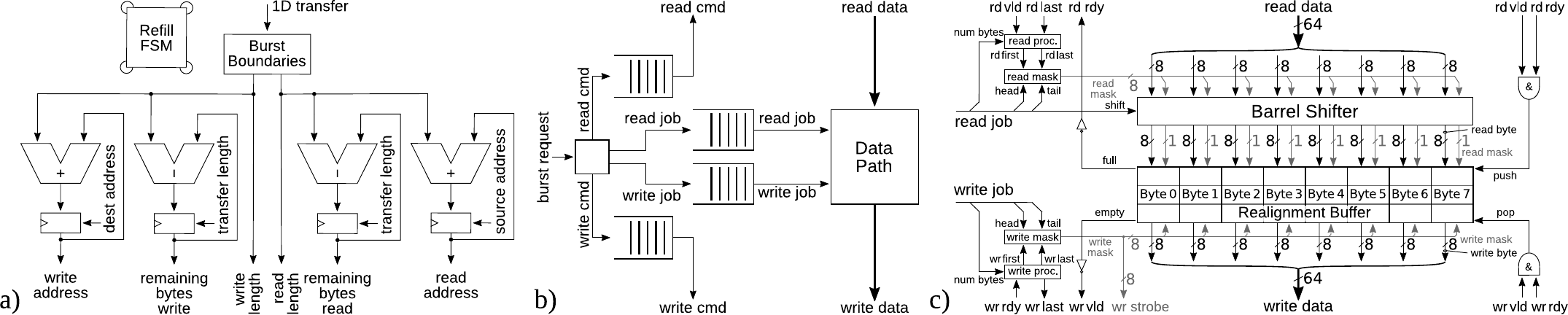}
    \caption{%
        Architecture of our \textbf{\acrshort{dma} engine}.
        (a) Burst reshaper.
        (b) Data mover.
        (c) Data path, drawn for \SI{64}{\bit} data width.
    }
    \label{fig:arch:dma}
\end{figure*}

Transferring large amounts of data at high bandwidth requires dedicated components for data movement called \emph{\glsunknown{dma} engines}.
Our \gls{dma} architecture is designed to be modular, dividing the unit into two parts: a system-specific frontend and a backend implementing the data movement within the on-chip interconnect.
We define a simple, yet well-defined interface uniting both parts:
a one-dimensional and contiguous memory block of arbitrary length, source, and destination address, called \definition{1D transfer}.
We chose this interface abstraction because 1D transfers map very well to burst-based transactions.
More complex transfers, such as multi-dimensional or strided accesses, are decomposed by the frontend into 1D transfers.
As the frontend is highly system-specific, we will not discuss it.

In the backend, the \definition{burst reshaper}, shown in \cref{fig:arch:dma}a, divides the arbitrary-length 1D transfers into protocol-compliant bursts (adhering to, e.g., address boundaries and maximum number of beats).
On arrival of a new 1D transfer, the burst converter loads length, source address, and destination address into internal registers.
The \definition{burst boundaries} process determines the number of bytes that can be requested in the next burst.
With this, the burst reshaper calculates the address of the next burst and the remaining bytes left in the 1D transfer.
Each protocol-compliant burst is then translated by the \emph{data mover} unit, shown in \cref{fig:arch:dma}b, into a read and a write command as well as a read and a write data job.
The commands are issued as beats on the command channels.
The data jobs are forwarded to the data path.
The \emph{data path}, shown in \cref{fig:arch:dma}c, receives read data beats, realigns the data to compensate for different byte offsets between the read and write data streams, and issues write data beats.
The data path consists of two independent processes.
The read process is realigns and buffers incoming data.
If a burst starts on an unaligned address, some leading bytes (``head'') in the first beat are invalid and are masked.
Similarly, a burst may end on an unaligned address, in which case some trailing bytes in the last beat (``tail'') need to be masked.
The write process drains data from the buffer as soon as it is available and masks it according to the destination address offset with the strobe signal of the write data channel.

\subsection{On-Chip Memory Controllers}%
\label{sec:arch:on-chip_mem_ctrl}

On-chip memories are an important class of endpoints for on-chip network transactions.
In this section, we describe two memory controllers through which standard single-port \gls{sram} macros can be connected to the on-chip network.

\subsubsection{Simplex Memory Controller}%
\label{sec:arch:on-chip_mem_ctrl:simplex}

\begin{figure}
    \centering%
    \newlength{\axitomemwidth}%
    \ifthesis%
      \setlength{\axitomemwidth}{.9\columnwidth}%
    \else%
      \setlength{\axitomemwidth}{.8\columnwidth}%
    \fi%
    \includegraphics[width=\axitomemwidth]{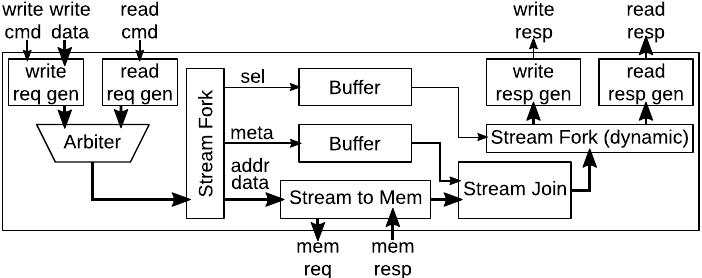}%
    \caption{%
      Architecture of our \textbf{simplex on-chip memory controller}, with the on-chip network slave port at the top and the memory master port at the bottom.
      The memory master port has the same data width as the network slave port.
    }%
    \label{fig:arch:axi_to_mem}%
\end{figure}

The architecture of our simplex on-chip memory controller is shown in \cref{fig:arch:axi_to_mem}.
\emph{Simplex} in this context means that the controller in each clock cycle can either read or write memory, as is natural for a single-port \gls{sram}.
The memory controller first translates read commands and write commands plus write data into memory commands.
An arbiter then forwards either a read or a write memory command per clock cycle.
This arbiter optionally takes \gls{qos} attributes of a command into account and can prioritize write beats, which cannot be interleaved due to \orderrule{3}, over read beats.
A stream fork unit splits address and data, which go to the memory interface, and meta data (e.g., the transaction ID), which are used by the memory controller to form responses in the network protocol.
A converter translates the address and data stream into memory interface signals (with stream flow control on the command and no handshaking on the response path).
The memory responses are then joined with meta data to form read or write responses, which are finally issued on the corresponding network response channel.

The simplex memory controller cannot achieve the full bidirectional bandwidth of the duplex on-chip network interface, which has separate channels for read and write data.
The duplex memory controller removes this limitation.

\subsubsection{Duplex Memory Controller}%
\label{sec:arch:on-chip_mem_ctrl:duplex}

\begin{figure}
    \centering%
    \newlength{\axitomembankedwidth}%
    \ifthesis%
      \setlength{\axitomembankedwidth}{.85\columnwidth}%
    \else%
      \setlength{\axitomembankedwidth}{.75\columnwidth}%
    \fi%
    \includegraphics[width=\axitomembankedwidth]{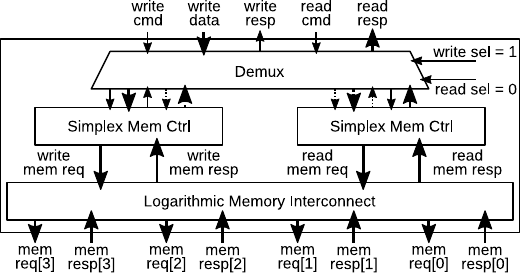}%
    \caption{%
      Architecture of our \textbf{duplex on-chip memory controller} with four address-interleaved memory master ports.
    }%
    \label{fig:arch:axi_to_mem_banked}%
\end{figure}

The architecture of our duplex memory controller is shown in \cref{fig:arch:axi_to_mem_banked}.
To saturate the read and write data channels of the on-chip network simultaneously (thus \definition{duplex}), this memory controller has at least two independent memory master ports as well as one simplex controller for writes and one for reads.
A network demultiplexer statically routes all writes through the left controller and all reads through the right controller.
The unused resources inside both simplex controllers are optimized away during synthesis.
A logarithmic memory interconnect then routes each command to one of the memory master ports, which are address-interleaved.

The duplex memory controller can fully saturate both the read and the write data channel of the on-chip network in the absence of conflicts on the memory ports.
However, irregular traffic (e.g., misaligned addresses, mixed wide and narrow beats) can give rise to a significant conflict rate.
To reduce conflicts, the \emph{banking factor} (i.e., the number of memory master ports per network slave port) can be increased to any integer higher than 2 (at the cost of more wide and shallow \gls{sram} macros when the memory capacity is to remain constant).

\section{Implementation Results}%
\label{sec:impl}

This section provides quantitative and asymptotic complexity results for our network modules.
These results are essential for architects to assess the feasibility and strike trade-offs in the design of on-chip networks.
\revhl{%
  Our findings are summarized in \cref{sec:impl:complexity}.
  Until there, this section discusses implementation results to derive the findings.
}

We implement the modules presented in \cref{sec:arch} in \gls{gf22} technology, using a ten-metal stack and eight track SLVT/LVT flip-well standard cells characterized at typical conditions (\SI{0.8}{\volt}, \SI{25}{\celsius}).
We synthesize with Synopsys DesignCompiler~2019.12 using topographical mode, so physical place-and-route constraints, dimensions, and delays are taken into account.
For the isolated implementation of the modules, each input is driven by a D-\gls{ff}, and each output drives a D-\gls{ff}.
Unless we vary it in the evaluation, we set the address and data width to \SI{64}{\bit} and the slave port ID width to \SI{6}{\bit}.
Before undergoing synthesis, all modules have been verified for protocol compliance in RTL simulation under extensive directed and constrained random verification tests.

\ifthesis%
  \subsection{Elementary\texorpdfstring{\,}{ }Components: Network\texorpdfstring{\,}{ }(De)Muxes}%
\else%
  \subsection{Elementary\texorpdfstring{\,}{ }Components: Network\texorpdfstring{\,}{ }(De)Multiplexers}%
\fi%
\label{sec:impl:mux_demux}

\subsubsection{Network Multiplexer}%
\label{sec:impl:mux}

\begin{figure}
    \centering
    \ifthesis%
      \includegraphics[width=.75\textwidth]{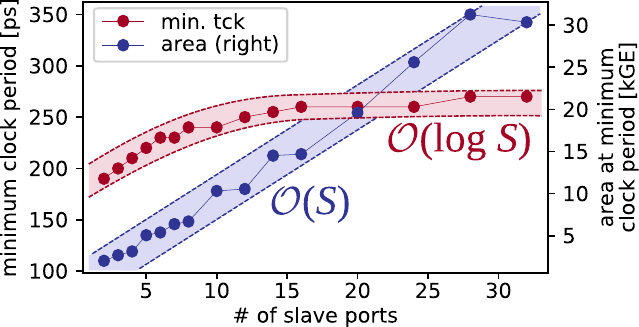}%
    \else%
      \includegraphics[height=.8in]{axi_mux__NumSlvPorts_max.pdf}%
    \fi%
    \label{fig:impl:mux:max}%
    \caption{%
        Minimum clock period and corresponding area of our network \textbf{multiplexer} in \acrshort{gf22} for \numrange{2}{32} slave ports and 6 ID bits.
    }%
    \label{fig:impl:mux}
\end{figure}

The critical path of the multiplexer goes through from a slave port command channel through the arbitration tree on its handshake signals and the multiplexers on its payload signals to a master port command channel.
For $S$ slave ports, it scales with \bigO{\log{S}} due to the logarithmic depth of the arbitration tree and the multiplexers.
The area scales \bigO{S} due to the linear area of the arbitration tree and the multiplexers.
The area is further linear in the ID width and the maximum number of write transactions due to the \gls{fifo} between write command and data channel, but this part is usually negligible.
\Cref{fig:impl:mux} shows the area and timing characteristics of our multiplexer:
for \numrange{2}{32} slave ports,
the critical path increases logarithmically from \SIrange{190}{270}{\pico\second},
and the area increases linearly from \SIrange{2}{30}{\kilo\GE}.

\subsubsection{Network Demultiplexer}%
\label{sec:impl:demux}

\begin{figure}
    \ifthesis%
      \centering
      \includegraphics[width=.75\textwidth]{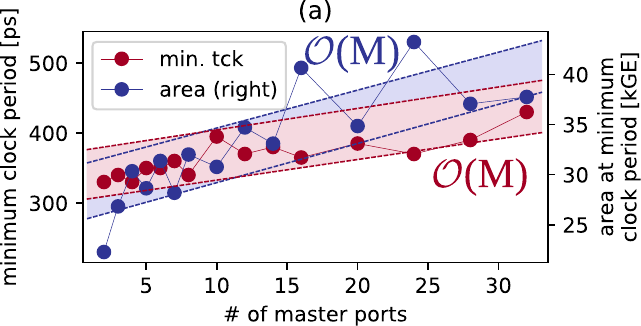}%
      \\
      \includegraphics[width=.75\textwidth]{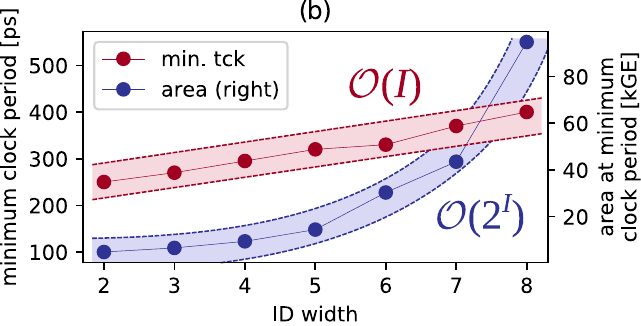}%
    \else%
      \maxsizebox{\columnwidth}{!}{%
        \includegraphics[height=0.9in]{axi_demux_IdWidth6_NumMstPorts_max.pdf}%
        \hspace*{.15em}%
        \includegraphics[height=0.9in]{axi_demux_NumMstPorts4_IdWidth_max.pdf}%
      }%
    \fi%
    \caption{%
      Minimum clock period and corresponding area of our network \textbf{demultiplexer} in \acrshort{gf22}:
      (a)~with \numrange{2}{32} master ports and 6 ID bits, and 
      (b)~with 4 master ports and \numrange{2}{8} ID bits.
    }%
    \label{fig:impl:demux}
\end{figure}

The critical path of the demultiplexer goes from a command channel at the slave port through ID lookup to a command channel on one of the master ports.
It scales with \bigO{M} as the channel demultiplexers grow linearly in area with the master ports and topographical synthesis takes the distance increase into account.
The area scales with \bigO{M} due to the linear area of the arbitration trees and the channel demultiplexers.
The ID width $I$ is critical for the demultiplexer: the area scales with \bigO{2\sp{I}} due to the exponential number of counters (one for every possible ID), and the critical path scales with \bigO{I} because every ID bit adds a multiplexer level in the indexing logic of the counters.
\Cref{fig:impl:demux} shows the area and timing characteristics of our demultiplexer:
For \numrange{2}{32} master ports and 6 ID bits (\cref{fig:impl:demux}a),
the critical path increases linearly from \SIrange{330}{430}{\pico\second},
and the area increases linearly from \SIrange{22}{38}{\kilo\GE}.
The curve is non-monotonic mainly in two points, where the synthesizer selects disproportionately strong and large buffers to reach the target frequency.
For 4 master ports and \numrange{2}{8} ID bits (\cref{fig:impl:demux}b),
the critical path increases linearly from \SIrange{250}{400}{\pico\second},
and the area increases exponentially from \SIrange{5}{95}{\kilo\GE}.
Depending on the ID width, the critical path can be significantly longer than in the multiplexer, so the demultiplexer will be the critical stage in a pipelined network junction.

\subsection{Network Junctions: Crossbars and Crosspoints}%
\label{sec:impl:junction}

\subsubsection{Crossbar}

\begin{figure}
    \ifthesis%
      \centering
      \includegraphics[width=.75\textwidth]{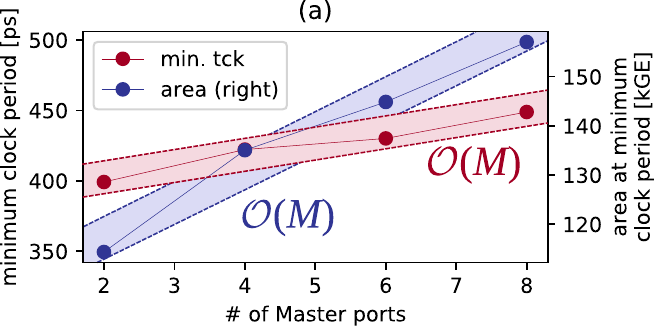}%
      \\
      \includegraphics[width=.75\textwidth]{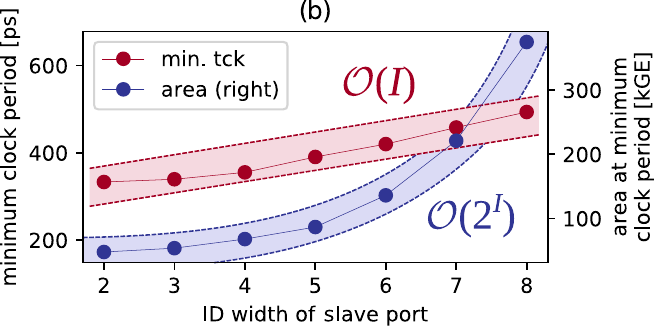}%
    \else%
      \maxsizebox{\columnwidth}{!}{%
        \includegraphics[height=0.9in]{results_xbar__axi_xbar_NumSlvPorts4PipelineStages0_NumMstPorts_max.pdf}%
        \hspace*{.05em}%
        \includegraphics[height=0.9in]{results_xbar__axi_xbar_PipelineStages0NumSlvPorts4NumMstPorts4_SlvPortIdWidth_max.pdf}%
      }%
    \fi%
    \caption{%
      Minimum clock period and corresponding area of our \textbf{crossbar} with 4 slave ports, fully connected and unpipelined, in \acrshort{gf22}:
      (a)~with \numrange{2}{8} master ports, 4 slave ports and 6 ID bits, and
      (b)~with 4 master ports and \numrange{2}{8} ID bits at the slave port.
    }%
    \label{fig:impl:xbar}
\end{figure}

For a fully-connected crossbar with $S$ slave ports, $M$ master ports and $I$ bits at the slave port, the critical path %
is dominated by the demultiplexer, thus scales with \bigO{M + I}.
The area is the sum of the area of the $S$ demultiplexers and $M$ multiplexers plus a small overhead for each slave port for address decoding and the error slave (when instantiated).
The area thus scales with \bigO{M S + 2\sp{I} S}.
\cref{fig:impl:xbar} shows the area and timing characteristics of a fully-connected, unpipelined instance of our crossbar:
For 4 slave ports, \numrange{2}{8} master ports and 6 ID bits (\cref{fig:impl:xbar}a), the critical path increases linearly from \SIrange{400}{450}{\pico\second},
and the area increases linearly from \SIrange{111}{156}{\kilo\GE}.
As was the case for the demultiplexer (\cref{sec:impl:demux}), the ID width of the slave ports has significant impact on the critical path and area of the crossbar.
For 4 master and 4 slave ports and \numrange{2}{8} ID bits (\cref{fig:impl:xbar}b), the critical path increases linearly from \SIrange{340}{460}{\pico\second}, and the area increases exponentially from \SIrange{42}{390}{\kilo\GE}.

\subsubsection{Crosspoint}%
\label{sec:impl:xp} 

\begin{figure}
    \ifthesis%
      \centering
      \includegraphics[width=.75\textwidth]{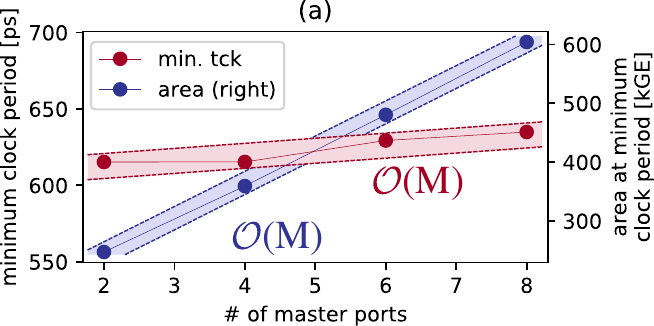}%
      \\
      \includegraphics[width=.75\textwidth]{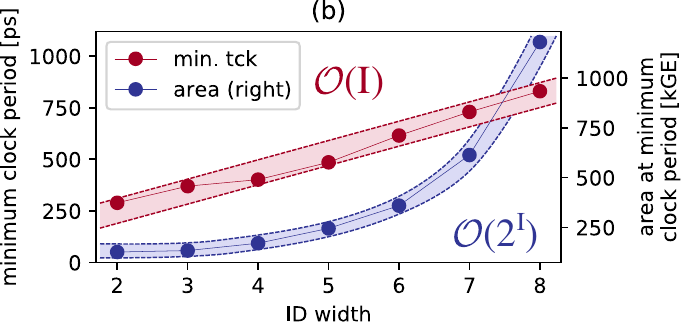}%
    \else%
      \maxsizebox{\columnwidth}{!}{%
        \includegraphics[height=0.9in]{results_xp__axi_xp_NumSlvPorts4SlvPortIdWidth6_NumMstPorts_max.pdf}%
        \hspace*{.05em}%
        \includegraphics[height=0.9in]{results_xp__axi_xp_NumSlvPorts4NumMstPorts4_SlvPortIdWidth_max.pdf}%
      }%
    \fi%
    \caption{%
      Minimum clock period and corresponding area of our \textbf{crosspoint} with 4 slave ports, fully connected and pipelined, in \acrshort{gf22}:
      (a)~with \numrange{2}{8} master ports, 4 slave ports and 6 ID bits, and
      (b)~with 4 master ports and \numrange{2}{8} ID bits at the ports.
    }%
    \label{fig:impl:xp}
\end{figure}

The critical path of a fully pipelined crosspoint goes from the internal pipeline register of a master port into the table of an ID remapper.
For $M$ master ports (\cref{fig:impl:xp}a), it scales with \bigO{M} from \SIrange{610}{630}{\nano\second} as topographical synthesis takes the area increase into account.
The area also scales with \bigO{M} but much more significantly from \SIrange{243}{587}{\kilo\GE} as the crossbar and the number of ID remappers scale linearly.
Regarding the ID width $I$, the crosspoint is dominated by the demultiplexer:
For \numrange{2}{8} ID bits in a $4 \times 4$ configuration (\cref{fig:impl:xp}b), the area scales with \bigO{2\sp{I}} from \SIrange{127}{1181}{\kilo\GE} and the critical path scales with \bigO{I} from \SIrange{290}{800}{\pico\second}.

\subsection{Concurrent Transactions: ID Width Converters}

\subsubsection{ID Remapper}%
\label{sec:impl:id_remapper}

\begin{figure}
  \ifthesis
    \centering
    \includegraphics[width=.75\textwidth]{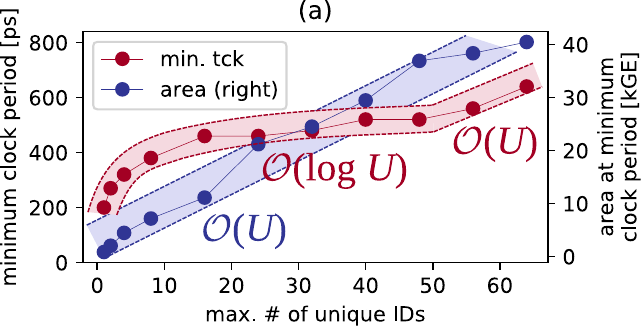}%
    \label{fig:impl:id_remapper:uniq_ids_max}%
    \\
    \includegraphics[width=.75\textwidth]{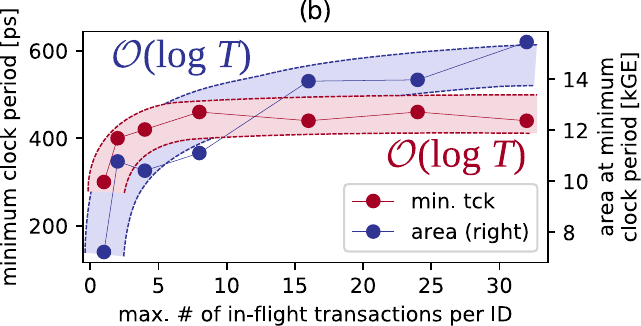}%
  \else
    \maxsizebox{\columnwidth}{!}{%
      \includegraphics[height=0.9in]{axi_id_remap_SlvPortIdWidth6MaxTxnsPerId8_SlvPortMaxUniqIds_max.pdf}%
      \label{fig:impl:id_remapper:uniq_ids_max}%
      \hspace*{.05em}%
      \includegraphics[height=0.9in]{axi_id_remap_SlvPortMaxUniqIds16MstPortIdWidth4_MaxTxnsPerId_max.pdf}%
    }%
  \fi
  \caption{%
      Minimum clock period and corresponding area of our \textbf{ID remapper} in \acrshort{gf22}:
      (a)~for \numrange{1}{64} concurrent unique IDs and 8 transactions per ID, and
      (b)~for 16 concurrent unique IDs and \numrange{1}{32} transactions per ID.
  }%
  \label{fig:impl:id_remapper}
\end{figure}

The critical path of our ID remapper goes from the input ID through the ID equality comparators in in the table, through a \gls{lzc} to determine the matching or the first free output ID, into a table counter entry.
For an input ID width of $I$, up to $U$ concurrent unique IDs (per direction), and up to $T$ transactions per ID, it scales with \bigO{\log{I} + \log{U} + \log{T}}.
The area is dominated by the tables, which have $U$ entries with $I + \log\sb{2}{T}$ bit each.
Additionally, the \glspl{lzc} have an area of \bigO{U \log{U}}.
The total area thus scales with \bigO{U (I + \log{T} + \log{U})}.
\Cref{fig:impl:id_remapper} shows the area and timing characteristics of our ID remapper:
For $U=$~\numrange{1}{64} concurrent unique IDs and $T=$~8 transactions per ID (\cref{fig:impl:id_remapper}a),
the critical path increases logarithmically from \SIrange{200}{520}{\pico\second} until $U = 48$ and then linearly to \SI{640}{\pico\second} for $U = 64$ as path delays due to the linearly growing table start to dominate.
The area increases linearly from \SIrange{1}{41}{\kilo\GE}.
The highest (rightmost) configuration can remap up to 512 transactions in both directions with up to 64 unique IDs concurrently, but the area and critical path costs are quite high.
In comparison, for $U = 16$ concurrent unique IDs and $T = $~\numrange{1}{32} transactions per ID (\cref{fig:impl:id_remapper}b),
the critical path increases logarithmically from \SIrange{300}{440}{\pico\second},
and the area increases logarithmically from \SIrange{7}{16}{\kilo\GE}.
Thus, the highest (rightmost) configuration can also remap up to 512 transactions but with only up to 16 unique IDs concurrently, at a $2.6\times$ lower area and $1.5\times$ shorter critical path.

\subsubsection{ID Serializer}%
\label{sec:impl:id_serializer}

\begin{figure}
  \ifthesis
    \centering
    \includegraphics[width=.75\textwidth]{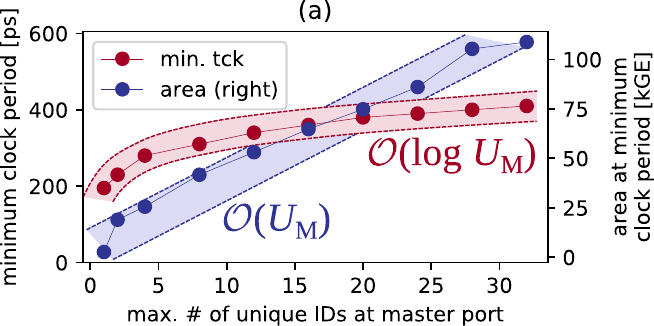}%
    \\
    \includegraphics[width=.75\textwidth]{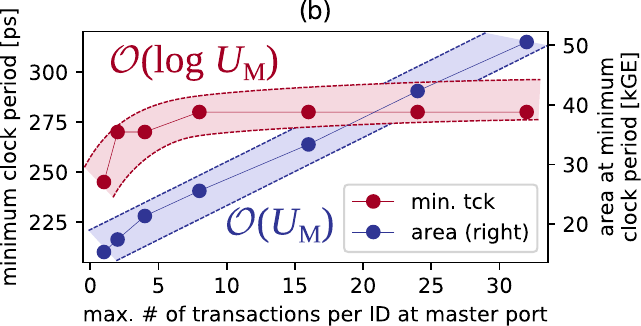}%
  \else
    \maxsizebox{\columnwidth}{!}{%
      \includegraphics[height=0.9in]{axi_id_serialize_SlvPortIdWidth6MstPortMaxTxnsPerId8_MstPortMaxUniqIds_max.pdf}%
      \hspace*{.05em}%
      \includegraphics[height=0.9in]{axi_id_serialize_SlvPortIdWidth6MstPortMaxUniqIds4_MstPortMaxTxnsPerId_max.pdf}%
    }%
  \fi
  \caption{%
      Minimum clock period and corresponding area of our \textbf{ID serializer} in \acrshort{gf22}:
      (a)~for \numrange{1}{32} IDs at the master port and 8 transactions per master port ID, and
      (b)~for 4 IDs and \numrange{1}{32} transactions per ID at the master port.
  }%
  \label{fig:impl:id_serializer}
\end{figure}

The critical path of the ID serializer goes through the demultiplexer, the push side of the ID \gls{fifo}, and the arbitration tree in the multiplexer.
For $U\sb{\text{M}}$ IDs at the master port and $T$ transactions per master port ID, it scales with \bigO{\log{U\sb{\text{M}}} + \log{T}}.
The area scales with \bigO{U\sb{\text{M}} + T} due to the linear area of all components in either $U\sb{\text{M}}$ or $T$.
\Cref{fig:impl:id_serializer} shows the area and timing characteristics of our serializer:
For $U\sb{\text{M}} = $~\numrange{1}{32} IDs at the master port and $T = 8$ transactions per master port ID (\cref{fig:impl:id_serializer}a),
the critical path increases logarithmically from \SIrange{195}{410}{\pico\second},
and the area increases linearly from \SIrange{2}{109}{\kilo\GE}.
Clearly, compressing a densely used ID space is expensive in terms of area.
This cost can be reduced by fixing $U\sb{\text{M}}$ at a low value and varying $T$:
For $U\sb{\text{M}} = 4$~IDs and $T = $~\numrange{1}{32} transactions per ID at the master port (\cref{fig:impl:id_serializer}b),
the critical path increases logarithmic from \SIrange{245}{280}{\pico\second},
and the area increases linearly from \SIrange{15}{51}{\kilo\GE}.
128 concurrent transactions (in both directions) could therefore be serialized with $U\sb{\text{M}} = 4, T = 32$ at $1.28\times$ less area and $1.29\times$ shorter critical path. %

\subsection{Data Width Converters}

\begin{figure}
    \centering
    \ifthesis
      \includegraphics[width=.75\textwidth]{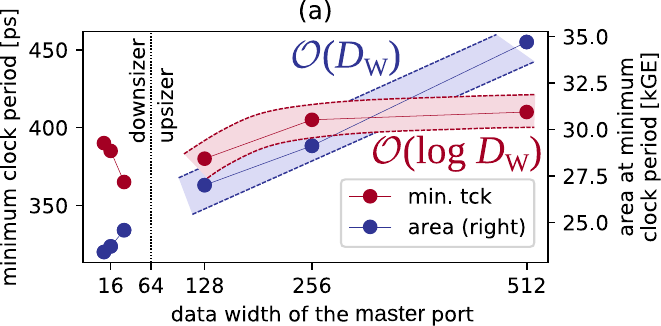}%
      \\
      \includegraphics[width=.75\textwidth]{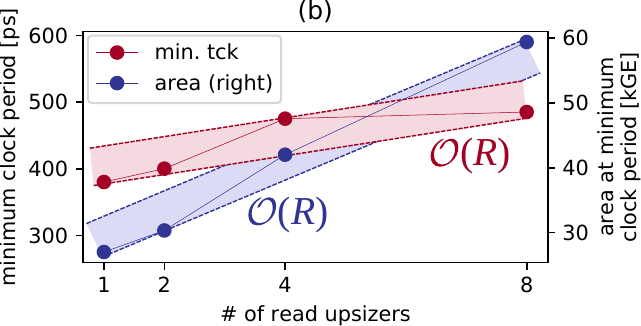}
    \else
      \maxsizebox{\columnwidth}{!}{%
        \includegraphics[height=0.9in]{results_dwc__axi_dwc_SlvPortDataWidth64MaxTxns1_MstPortDataWidth_max.pdf}%
        \hspace*{.05em}%
        \includegraphics[height=0.9in]{results_dwc__axi_dwc_SlvPortDataWidth64MstPortDataWidth128_MaxTxns_max.pdf}
      }%
    \fi
    \caption{Minimum clock period and corresponding area of: (a) our \textbf{data downsizer and upsizer}, considering a slave port data width of \SI{64}{\bit} and a master port data width from \SIrange{8}{512}{\bit}, and (b) our \textbf{data upsizer}, considering a slave port data width of \SI{64}{\bit}, a master port data width of \SI{128}{\bit}, and \numrange{1}{8} read upsizers.}
    \label{fig:impl:dwc}
\end{figure}

For our data downsizer between a wide slave port of width $D\sb{\text{W}}$ and a narrow master port of width $D\sb{\text{N}}$, the critical path goes through the data selection and steering logic, scaling logarithmically with the downsize ratio $\bigO{\log{\left(D\sb{\text{W}}/D\sb{\text{N}}\right)}}$. 
The area is $\bigO{D\sb{\text{N}} D\sb{\text{W}}}$, the first term accounting for the multiplexing logic for data selection and steering, and the second accounting for the registers that hold a wide beat for data packing on the write data channel. 
\Cref{fig:impl:dwc}a (left side) shows the area and timing characteristics of our downsizer: for a slave port of width \num{64} bits and a master port of width \numrange{8}{32} bits, the critical path decreases with increasing width of the master port (and decreasing downsize ratio), from \SIrange{390}{365}{\pico\second}, while the area grows linearly from \SIrange{23}{25}{\kilo\GE}.

For the data upsizer between a narrow slave port of width $D\sb{\text{N}}$ and a wide master port of width $D\sb{\text{W}}$, the critical path goes through the data selection logic and the round-robin arbiter, scaling linearly with the number of read upsizers $R$ and logarithmically with the upsize ratio, $\bigO{R \log\left({D\sb{\text{W}}/D\sb{\text{N}}}\right)}$.
The area of the upsizer scales with \bigO{R D\sb{\text{N}} D\sb{\text{W}}}, compounding the effect of the multiplexing logic for data selection and steering, $D\sb{\text{N}}$, and of the $R$ $D\sb{\text{W}}$-bit registers holding wide beats for data serialization on the read data channel.
\Cref{fig:impl:dwc}a (right side) shows the area and timing characteristics of our upsizer: for a slave port of width \num{64} bits and a master port of width \numrange{128}{512} bits, the critical path increases with the increasing upsize ratio, from \SIrange{380}{405}{\pico\second}, while the area increases from \SIrange{27}{35}{\kilo\GE}.
\Cref{fig:impl:dwc}b shows the area and timing characteristics of the data upsizer from \SIrange{64}{128}{bits}, for \numrange{1}{8} read upsizers.
These have an important effect on the area and critical path of the upsizer. 
The critical path of the upsizer increases linearly from \SIrange{380}{485}{\pico\second}, while the area increases from \SIrange{27}{59}{\kilo\GE}.

\color{colorrevhl}%
\subsection{Clock Domain Crossing}

The area of the \gls{cdc} scales linearly with the address, data, and ID widths.
For \SI{64}{\bit} address and data width, \SI{6}{\bit} ID width, and a slave port clock frequency of \SI{1}{\giga\hertz}, the \gls{cdc} area is \SI{27}{\kilo\GE} for master port clock frequencies from \SIrange{0.1}{2}{\giga\hertz}.
Above that, the area increases exponentially but only up to \SI{31}{\kilo\GE} for \SI{5.5}{\giga\hertz} at the master port.
\color{black}%

\subsection{Data Streaming: DMA Engine}

\begin{figure}
  \ifthesis
    \centering
    \includegraphics[width=.75\textwidth]{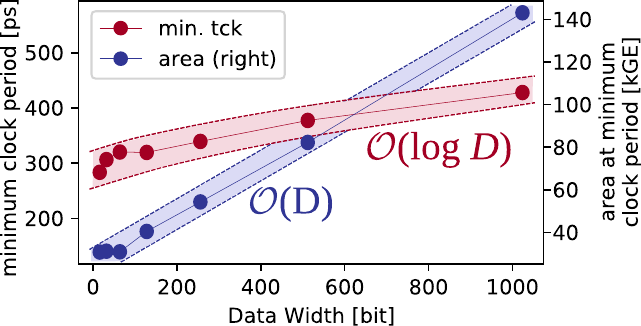}%
    \\
    \includegraphics[width=.75\textwidth]{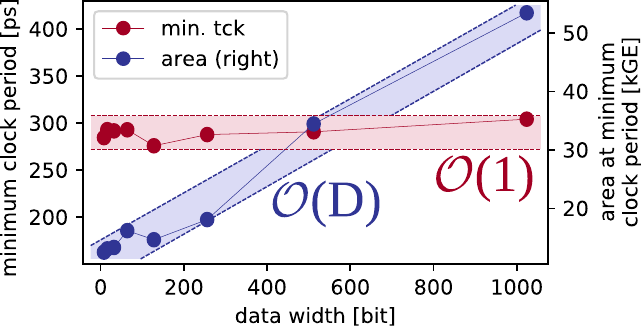}%
  \else
    \maxsizebox{\columnwidth}{!}{%
      \includegraphics[height=0.9in]{axi_dma_backend_BufferDepth3_DataWidth_max.pdf}%
      \hspace*{.05em}%
      \includegraphics[height=.9in]{axi_to_mem_NumBanks1BufDepth1_DataWidth_max.pdf}%
    }%
  \fi
  \caption{Minimum clock period and corresponding area in \acrshort{gf22} of (a) our \textbf{\acrshort{dma} engine} for \SIrange{16}{1024}{\bit} data width, and (b) our \textbf{simplex on-chip memory controller} for \SIrange{8}{1024}{\bit} data width.}%
  \label{fig:impl:dma_and_simplex_mem}
\end{figure}

The area of the \gls{dma} engine scales with \bigO{D}, where $D$ is the data width, due to the linearly growing alignment buffer.
The critical path is dominated by the barrel shifter, which scales with \bigO{\log{D}}.
For \SIrange{16}{1024}{\bit} data width (\cref{fig:impl:dma_and_simplex_mem}), the critical path increases logarithmically from \SIrange{290}{400}{\pico\second} and the area increases linearly from \SIrange{25}{141}{\kilo\GE}.
As the \gls{dma} engine uses the same ID for all transactions, the ID width affects neither area nor critical path.

\subsection{On-Chip Memory Controllers}
\label{sec:impl:mem_controller}

\subsubsection{Simplex Memory Controller}%
\label{sec:impl:mem_controller:simple}

For a simplex on-chip memory controller with a data width of $D$, the critical path is constant and found between the command slave channels and the memory master port.
The critical path does not depend on $D$ as the transformation of commands does not depend on the data width.
\Cref{fig:impl:dma_and_simplex_mem}b shows the area and timing characteristics:
The area scales linearly with \bigO{D} from \SIrange{13}{53}{\kilo\GE}; this linear dependency is caused by the dominant read response buffers needed for response path decoupling.
The critical path remains roughly constant around \SI{290}{\pico\second}.
The ID width has no impact on the critical path, as the simplex controller handles all commands in order and only buffers the ID for the response.
The area scales with \bigO{I} due to these buffers.

\subsubsection{Duplex Memory Controller}%
\label{sec:impl:mem_controller:banked}

\begin{figure}
  \ifthesis
    \centering
    \includegraphics[width=.75\textwidth]{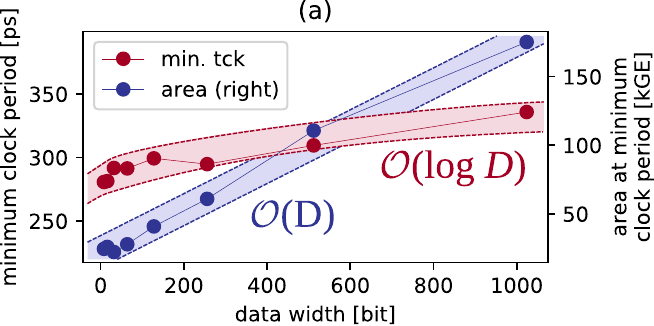}%
    \\
    \includegraphics[width=.75\textwidth]{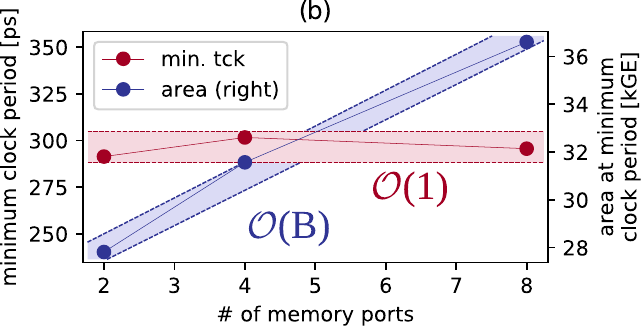}%
  \else
    \maxsizebox{\columnwidth}{!}{%
      \includegraphics[height=0.9in]{results_to_mem_banked__axi_to_mem_banked_NumBanks2BufDepth1_DataWidth_max.pdf}%
      \hspace*{.05em}%
      \includegraphics[height=0.9in]{results_to_mem_banked__axi_to_mem_banked_DataWidth64BufDepth1_NumBanks_max.pdf}%
    }%
  \fi
  \caption{%
      Minimum clock period and corresponding area of our \textbf{duplex on-chip memory controller} in \acrshort{gf22}:
      (a)~for \SIrange{8}{1024}{\bit} data width and two memory master ports, and
      (b)~for \SI{64}{\bit} data width and \numrange{1}{8} memory master ports.
  }%
  \label{fig:impl:to_mem_banked}
\end{figure}

The critical path of the duplex controller goes from the slave port command channels through the demultiplexer, one simplex memory controller, and the logarithmic memory interconnect to a memory command port.
For a data width of $D$ and $B$ memory master ports, it scales with \bigO{\log{D}}.
The area is composed of the demultiplexer, the two simplex memory controllers, and the logarithmic interconnect, and thus scales with \bigO{B + D}.
\Cref{fig:impl:to_mem_banked} shows the area and timing characteristics of our duplex memory controller:
For $D =$~\numrange{8}{1024} bit data width and $B = 2$ memory master ports (\cref{fig:impl:to_mem_banked}a), the critical path increases logarithmically from \SIrange{280}{330}{\pico\second}, and the area increases linearly from \SIrange{20}{175}{\kilo\GE}.
For $D = 64$ bit data width and $B =$~\numrange{2}{8} memory master ports (\cref{fig:impl:to_mem_banked}b), the critical path stays constant around \SI{300}{\pico\second} and the area scales with \bigO{B} from \SIrange{28}{34}{\kilo\GE}.
Regarding the ID width $I$, the complexity is defined by the demultiplexer.

\subsection{Complexity Overview and Summary}%
\label{sec:impl:complexity}

\begin{table}
    \begin{center}
        \setlength{\tabcolsep}{2pt}
        \maxsizebox{\columnwidth}{!}{
        \begin{tabular}{@{} l c c @{}}
            \toprule
            & \thead[c]{Critical Path} & \thead[c]{Area} \\
            \midrule
            Multiplexer & \bigO{\log{S}} & \bigO{S} \\
            Demultiplexer & \bigO{M + I} & \bigO{M + 2\sp{I}} \\
            Crossbar & \bigO{M + I} & \bigO{M S + 2\sp{I} S} \\
            Crosspoint & \bigO{M + I} & \bigO{M + 2\sp{I}} \\
            ID Remapper & \bigO{\log{I} + \log{U} + \log{T}} & \bigO{U (I + \log{T} + \log{U})} \\
            ID Serializer & \bigO{\log{U\sb{\text{M}}} + \log{T}} & \bigO{U\sb{\text{M}} + T} \\
            Data Upsizer & \bigO{R \log\left(D\sb{\text{W}}/D\sb{\text{N}}\right)} & \bigO{RD\sb{\text{W}}D\sb{\text{N}}} \\
            Data Downsizer & \bigO{\log\left(D\sb{\text{W}}/D\sb{\text{N}}\right)} & \bigO{D\sb{\text{W}}D\sb{\text{N}}} \\
            \acrshort{dma} Engine & \bigO{\log{D}} & \bigO{D} \\
            Simplex Mem.\ Ctrl.\ & \bigO{1} & \bigO{D} \\
            Duplex Mem.\ Ctrl.\ & \bigO{\log{D} + \log{B} + I} & \bigO{D + B + 2\sp{I}} \\
            \bottomrule
        \end{tabular}
        }
    \end{center}
    \footnotesize{%
      Legend:
      $M$~=~number of master ports;
      $S$~=~number of slave ports.
      $D$~=~data width;
      $D\sb{\text{W}}$~=~data width of the wide interface;
      $D\sb{\text{N}}$~=~data width of the narrow interface;
      $I$~=~ID width;
      $U$~=~concurrent unique IDs;
      $U\sb{\text{M}}$~=~concurrent unique IDs at the master port;
      $T$~=~concurrent transactions per ID.
      $B$~=~number of memory master ports.
      $R$~=~number of read upsizers.
    }
    \caption{Overview of the complexity of our network modules.}%
    \label{tab:impl:complexity}
\end{table}

An overview of the asymptotic complexity of our network modules is shown in \cref{tab:impl:complexity}.
The critical path of all modules scales at worst linearly in their parameters, for most modules and parameters even logarithmically.
As the absolute results of the minimum clock period show, the critical path of all modules remains below \SI{500}{\pico\second} post-topographical-synthesis in the large design space we evaluated.
This shows our modules are suited for a wide range of target frequencies and bandwidths, up to \SI{2}{\giga\hertz}.
When even higher frequencies are required, most modules can be parametrized to have a critical path below \SI{330}{\pico\second}, which would allow to clock them up to \SI{3}{\giga\hertz}.
The area of most modules scales linearly in their parameters, with the notable exception of the ID width, which causes an exponential growth of the demultiplexer and all modules containing it.
As the absolute results show, most modules fit a few tens of \si{\kilo\GE} when not pushed to the highest possible clock frequency and parametrization.
Even more complex modules, such as a $4 \times 4$ crossbar with up to 256 independent concurrent transactions, fit in a modest \SI{100}{\kilo\GE} when clocked at \SI{2.5}{\giga\hertz}.

Power is another important aspect of on-chip networks.
This paper focuses on architecture, performance, and area, but our platform supports all state-of-the-art power optimization techniques (e.g., architectural power gating).
Even for complex and high-performance instances such as the mentioned \SI{100}{\kilo\GE} crossbar, the power consumption is in the order of just \SI{35}{\milli\watt} under full load at \SI{2.5}{\giga\hertz}, which would typically be less than \SI{10}{\percent} of the power consumed by processor cores operating at a similar frequency; for instance, an ARM Cortex-A72 in a \SI{16}{\nano\meter} FinFET technology at \SI{2.5}{\giga\hertz} consumes around \SI{1}{\watt}~\cite{frumusanu2016cortexa72}.

While module-wise results are important to show the complexity and trade-offs in the microarchitecture of our on-chip communication platform, they of course cannot show the full picture of a real on-chip network.
In the next section, we analyze a full on-chip network. %

\ifpspin
\else%
  \section{Full-System Case Study}%
\fi%
\label{sec:sys}

In this section, we design, implement, and evaluate the on-chip network{} of a many-core floating-point accelerator%
\ifpspin
\else%
  , using the modules presented in this paper.
\fi%
We use the technology and synthesis flow described in \cref{sec:impl} and additionally implement the networks with Cadence~Innovus~19.1.

\begin{figure}
    \centering
    \includegraphics[width=.75\columnwidth]{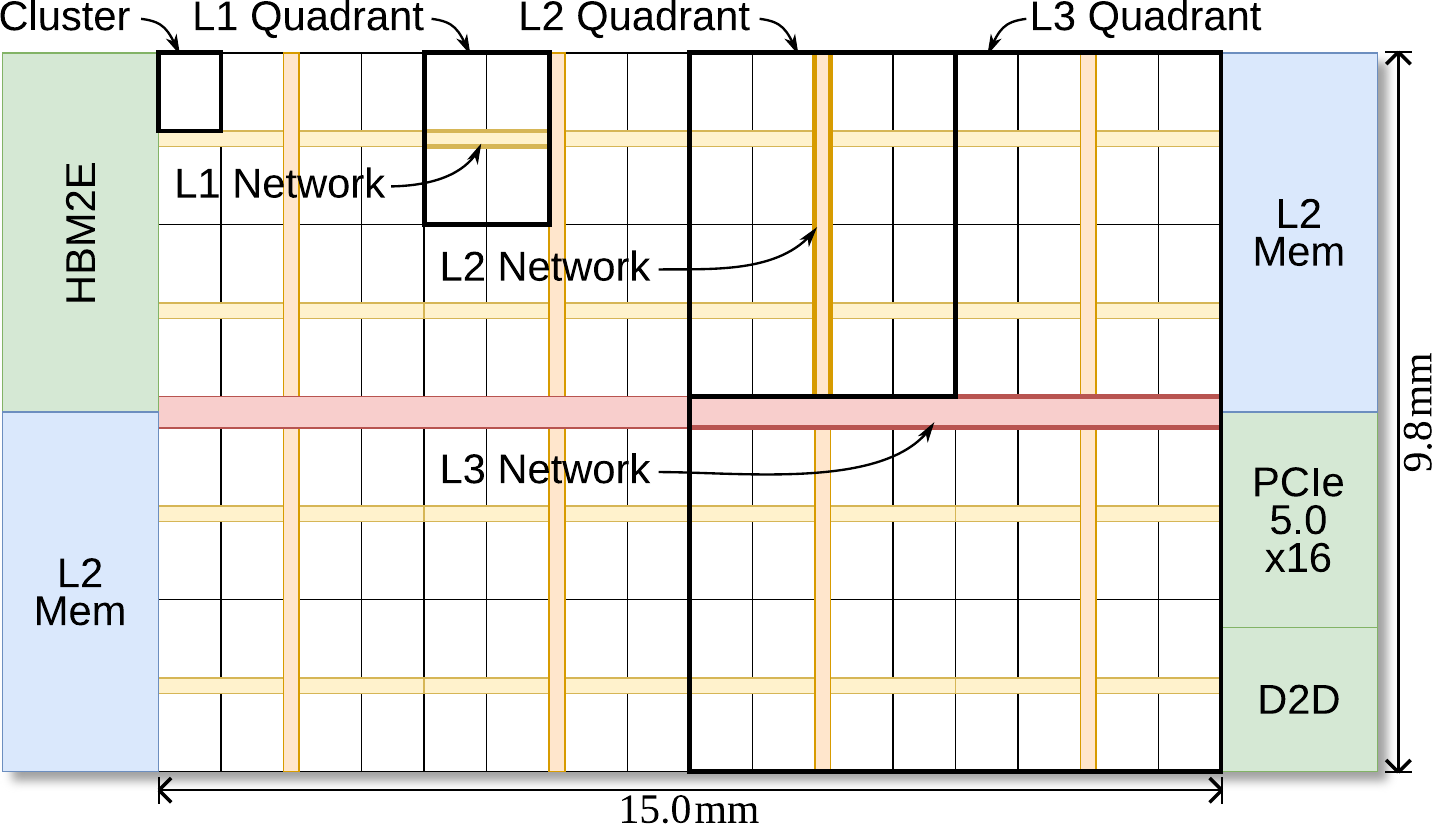}
    \caption{Conceptual floorplan of one Manticore~\cite{manticore} chiplet die.}%
    \label{fig:sys:manticore:floorplan}
\end{figure}

The \emph{Manticore} architecture~\cite{manticore} is a state-of-the-art manycore processor for high-performance, high-efficiency, data-parallel floating-point computing.
A Manticore accelerator consists of four chiplet dies on an interposer.
Each chiplet, shown in \cref{fig:sys:manticore:floorplan}, contains 1024 cores grouped in 128 clusters, one \SI{8}{\gibi\byte} HBM2E controller and PHY, \SI{27}{\mebi\byte} L2 memory, one \gls{pcie}~5.0~x16 controller and PHY, and three \gls{d2d} PHYs to the other chiplets.
Each cluster contains eight small 32-bit integer RISC-V cores, each controlling a large double-precision \gls{fpu}, and \SI{128}{\kibi\byte} L1 memory organized in 32 \gls{sram} banks.
As primary means for moving data into and out of L1, each cluster contains two of our \glsknown{dma} engines~(\cref{sec:arch:dma_engine}, one for reads and one for writes), which are attached to the L1 memory and control a 512-bit-wide master port.
\Gls{dma} engines in other clusters can access the L1 memory through an additional 512-bit-wide slave port.
Each cluster has a 64-bit master port to let its cores access external memory and a 64-bit slave port to let cores in other clusters access its L1 memory.
Four clusters form an L1 quadrant, four L1 quadrants form an L2 quadrant, four L2 quadrants form an L3 quadrant, and two L3 quadrants form a chiplet.
Manticore has been introduced in~\cite{manticore} without disclosing its on-chip network.
\revhl{In \ifpspin\else the remainder of this section\fi, we describe the design, implementation, and performance of Manticore's on-chip network.}

\ifpspin
\else
  \subsection{Network Design}
\fi

\begin{figure}
    \centering
    \includegraphics[width=.95\columnwidth]{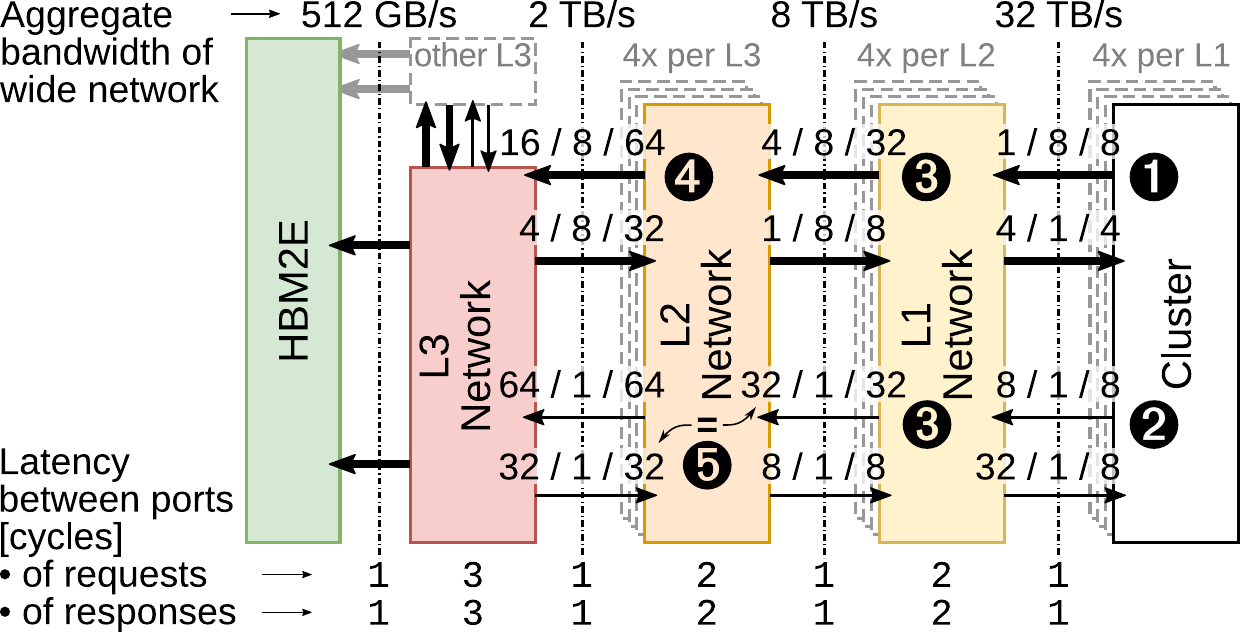}
    \caption{%
      Manticore's on-chip network.
      Each arrow represents a \revhl{bundle, with the arrow head pointing in the direction of the command channels}.
      Fat arrows mean \SI{512}{\bit} data width, thin arrows \SI{64}{\bit}.
      Numbers above arrows indicate maximum transaction concurrency in the form \textit{unique IDs} / \textit{transactions per ID} / \textit{total transactions per bundle} (reads and writes separate).
    }%
    \label{fig:sys:manticore:network}
\end{figure}

\renewcommand{\goal}[1]{\revhl{\textbf{(D#1)}}}
Manticore's on-chip network is designed with four main goals:
\goal{1}~High bandwidth between units within the same quadrant for effective local data sharing.
\goal{2}~High bandwidth between the chiplet-level \glsknownpl{io} (i.e., HBM2E, PCIe, \gls{d2d}) and any cluster for effective data input and output.
\goal{3}~Low latency between any two cores for efficient concurrency.
\goal{4}~Minimal interference between the wide bursts of the \gls{dma} engine and the word-wise accesses of the cores for maximum network utilization.
The network, shown in \cref{fig:sys:manticore:network}, has the following properties to meet these goals:
(1)~Physically separate networks for traffic by \gls{dma} engines and cores to meet \goal{4}.
(2)~Tree topology to meet \goal{2--3}.
(3)~Fully-connected crossbars within each quadrant to meet \goal{1}.
\revhl{(4)~The same data width and frequency for each bundle throughout the \gls{dma} network, to meet \goal{2}.}
The clock frequency of the entire network \revhl{and all its endpoints (i.e., cores, \gls{dma} engines, L2 memory controllers, and HBM2E controller)} is \SI{1}{\giga\hertz}.
The data width of the \gls{dma} network is set to \SI{512}{\bit}, which corresponds to one of the four ports into the HBM2E controller.
Therefore, saturating the full HBM2E bandwidth requires concurrent transactions from only four \gls{dma} engines in different L2 quadrants.
The data width of the core network is set to \SI{64}{\bit}, which is native for the load/store unit of a core.

The concurrency of transactions is another important aspect of the network design.
The numbers above an arrow in \cref{fig:sys:manticore:network} define the number of concurrent unique IDs, transactions per ID, and total transactions per bundle (reads and writes separate), respectively.
ID width converters are placed in the network where required to reduce the ID width.
Starting at the cluster, each \gls{dma} engine is in-order (thus has a single ID) and can have up to 8 outstanding transactions~\negcircnum{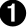}.
Transactions by the 8 cores in the cluster are independent, and each core can have at most 1 outstanding transaction~\negcircnum{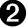}.
The L1 network maintains the independence of all \gls{dma} and core transactions, and the number of unique IDs expands accordingly, as do the total transactions~\negcircnum{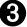}.
The L2 network maintains the independence of \gls{dma} transactions but limits their total below the sum of the incoming ports \revhl{with ID remappers}~\negcircnum{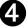}.
The reason is that %
the maximum roundtrip latency at this level is 60 cycles,
so a higher number of concurrent transactions would not increase bandwidth or utilization.
The concurrency on downlinks is generally constrained \revhl{with ID remappers} to match that of an uplink into the lower network, e.g.,~\negcircnum{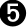}.
This means each network level can handle transactions from the uplink slave port in the same way as transactions from downlink slave ports.

\ifpspin
\else
  \subsection{Network Microarchitecture and Implementation}
\fi

\begin{figure}
    \includegraphics[width=\columnwidth]{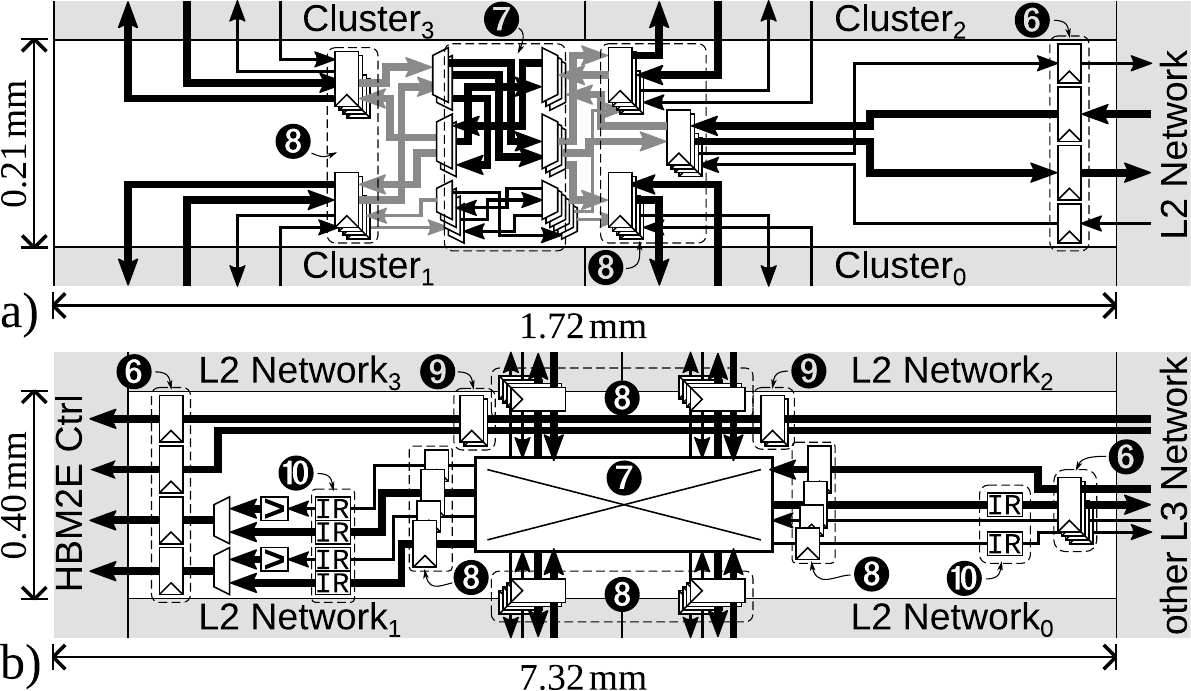}
    \caption{Microarchitecture and dimensions of Manticore's on-chip network.
    (a) L1 network.
    (b) L3 network.
    Only select connections are drawn for reasons of lucidity.
    }%
    \label{fig:sys:manticore:bds}
\end{figure}

The microarchitecture and physical dimensions of one L1 and L3 network are shown in \cref{fig:sys:manticore:bds}.
(The L2 network is very similar to the L1 and omitted for brevity.)
For the L1 network, the downlink ports are in the left third of each cluster, close to the cluster's memory and internal interconnect, and the uplink port is in the middle of the narrow side.
For the L2 and L3 network, the downlink ports are at one quarter of the wide side (determined by the lower network level), and the uplink port is in the middle of the narrow side.
To isolate the timing closure of individual network levels, we cut all paths at the uplink ports~\negcircnum{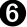}.
Correspondingly, all downlink inputs are driven by \glspl{ff} and all downlink outputs drive \glspl{ff}.
There are two central challenges in the physical implementation of the networks.
First, the extremely wide aspect ratio: while one wide dimension is determined by the side length quadrant, the other dimension should be as narrow as possible to minimize the area of the network.
Second, routing and wire congestion: each of the five interfaces has ca.\ \num{3300} separate wires, and each network level is fully connected.
Routing the wires of a single interface horizontally occupies a height of ca.\ \SI{100}{\micro\meter} on all three metal layers available for inter-cell horizontal routing.
To mitigate congestion, %
the crossbar, with its fanout of wires between demultiplexers and multiplexers, should be placed and routed
as compact as possible~\negcircnum{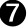}.
The crossbar nonetheless incurs a significant combinational delay. %
To accommodate this despite the long distances due to the extreme aspect ratio,
we insert registers %
around the crossbar~\negcircnum{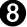}.
In contrast to pipelining inside the crossbar, much fewer registers are required, which again benefits the compact layout of the crossbar.
In the L3 network (\cref{fig:sys:manticore:bds}b), pairs of L2 networks share one port on the HBM2E controller.
Cores on the narrow network access the wide HBM ports through data width converters.
Because the HBM2E controller is located on the left side of the chiplet, the left L3 network simply feeds two connections from the right L3 network through pipeline registers to the controller~\negcircnum{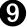}.
ID remappers are used to reduce ID widths according to the concurrency design~\negcircnum{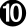}.

\begin{table}
    \begin{center}
        \setlength{\tabcolsep}{4pt}
        \begin{tabular}{ l l r r r r }
            \toprule
            {} & \thead{Unit} & \thead{L1} & \thead{L2} & \thead{L3} & \thead[l]{Entire\\Network} \\
            \midrule
            Clock Frequency & [\si{\giga\hertz}] & 1.00 & 1.00 & 1.00 & 1.00 \\
            Routing Density* & [\si{\percent}] & 59.6 & 49.6 & 45.7 & --- \\
            Area per Inst.\ & [\si{\milli\meter\squared}] & 0.41 & 1.40 & 2.99 & 30.43 \\
            Power per Inst.\ & [\si{\milli\watt}] & 8.1 & 12.8 & 17.2 & 396.0 \\
            \# Insts. per Chiplet & \vphantom{[\si{\milli\meter\squared}]} & 32 & 8 & 2 & 1 \\
            Area per Chiplet & [\si{\milli\meter\squared}] & 13.21 & 11.23 & 5.98 & 30.43 \\
            Area per Chiplet${}\sp{\dagger}$ & [\%] & 9.05 & 7.69 & 4.10 & 20.84 \\
            Area per Core+FPU & [\si{\micro\meter\squared}] & \num{12900} & \num{10970} & \num{5840} & \num{29710} \\
            Power per Chiplet & [\si{\milli\watt}] & 259.2 & 102.4 & 34.4 & 396.0 \\
            \bottomrule
        \end{tabular}
    \end{center}
    \footnotesize{%
        *Routing density along wider dimension (i.e., where routing is denser).\\
        ${}\sp{\dagger}$Relative to chiplet area without \acrshort{io} controllers and PHYs.
    }
    \caption{Implementation results of Manticore's on-chip network.}%
    \label{tab:manticore:results}
\end{table}

The implementation results of Manticore's on-chip network are listed in \cref{tab:manticore:results}.
We have been able to close timing and \gls{drc} of the entire network after place and route at \SI{1}{\giga\hertz}.
For this, we first loosely constrained the narrow dimension to determine the required number of pipeline registers around the crossbar, then we reduced the narrow dimension until the design could no longer be routed without failing timing or \gls{drc}.
As the high routing densities show, the area of each network level is mainly determined by the available routing channels.
The total area of the network is \SI{30.43}{\milli\meter\squared}, which is \SI{20.84}{\percent} of the chiplet area without \gls{io} controllers and PHYs.
Put differently, Manticore's entire high-bandwidth, low-latency, hierarchical on-chip network requires \SI{29710}{\micro\meter\squared} per core.
This is merely about the same area as one core (without any cache) and \gls{fpu}, which are %
highly area-efficient.
The total power of the network is \SI{396}{\milli\watt}, which means only \SI{0.4}{\milli\watt} per core (or less than \SI{5}{\percent} extra power per core).

\ifpspin
\else
  \subsection{Network Performance}
\fi

We characterize the performance of two fundamental \gls{nn} layers based on RTL simulations and analytical calculations, with a focus on the impact of the on-chip network.
The two layers, a convolutional layer and a fully-connected layer, together account for \SIrange{95}{99}{\percent} of the \glspl{flop} in \gls{mlt}.
The following is a condensed description of the \gls{nn} implementation on Manticore, a comprehensive description is available as online supplement\,\footnote{\url{https://arxiv.org/pdf/2104.08009.pdf}}.

\begin{table}
    \begin{center}
        \setlength{\tabcolsep}{3pt}
        \begin{tabular}{ l l r r r r }
            \toprule
            \thead{Figure} & \thead{Unit} & \multicolumn{3}{c}{\textbf{Convolution}} & \multicolumn{1}{c}{\textbf{Fully}} \\
            {} & {} & base & stacked & pipe'd & \multicolumn{1}{c}{\textbf{Connected}} \\
            \midrule
            Op.\ Intensity & [\si{\dpflop\per\byte}] & 2.2\hphantom{*} & 15.9\hphantom{${}\sp{\dagger}$} & 15.9\hphantom{${}\sp{\dagger}$} & 7.9\hphantom{${}\sp{\dagger}$} \\
            HBM BW & [\si{\giga\byte\per\second}] & 262* & 98\hphantom{${}\sp{\dagger}$} & 6\hphantom{${}\sp{\dagger}$} & 222\hphantom{${}\sp{\dagger}$} \\
            L3 Agg.\ BW & [\si{\giga\byte\per\second}] & 262\hphantom{*} & 98\hphantom{${}\sp{\dagger}$} & 6\hphantom{${}\sp{\dagger}$} & 222\hphantom{${}\sp{\dagger}$} \\
            L2 Agg.\ BW & [\si{\giga\byte\per\second}] & 262\hphantom{*} & 98\hphantom{${}\sp{\dagger}$} & 25\hphantom{${}\sp{\dagger}$} & 222\hphantom{${}\sp{\dagger}$} \\
            L1 Agg.\ BW & [\si{\giga\byte\per\second}] & 262\hphantom{*} & 98\hphantom{${}\sp{\dagger}$} & 98\hphantom{${}\sp{\dagger}$} & 222\hphantom{${}\sp{\dagger}$} \\
            Performance & [\si{\giga\dpflop\per\second}] & 571\hphantom{*} & 1638${}\sp{\dagger}$ & 1638${}\sp{\dagger}$ & 1638${}\sp{\dagger}$ \\
            \bottomrule
        \end{tabular}
    \end{center}
    *Of which \SI{256}{\giga\byte\per\second} are on the read channel, which is its maximum.\\
    ${}\sp{\dagger}$This corresponds to an \acrshort{fpu} utilization of ca.\ \SI{80}{\percent}, which is the maximum all 8 \glspl{fpu} in a cluster can sustain for real kernels.
    \caption{%
      Performance of Manticore for different \acrshort{nn} layer implementations.
    }%
    \label{tbl:sys:manticore:perf}
\end{table}

A convolutional \gls{nn} layer transforms a 3D input volume (aka ``feature map'') to a 3D output volume through convolutions with filter kernels.
More precisely, each input volume with dimensions $W\sb{I} \times W\sb{I} \times D\sb{I}$ is padded with $P$ zeros in the spatial dimensions (resulting in $(W\sb{I} + 2 P) \times (W\sb{I} + 2 P)$) and then convolved with $K$ filter kernels with dimension $F \times F \times D\sb{I}$ in a stride $S$ to produce an output volume with dimensions $W\sb{O} \times W\sb{O} \times D\sb{O}$, where $W\sb{O} = (W\sb{I} + 2 P - F) / S + 1$ and $D\sb{O} = K$.
We set $W\sb{I} = 32$, $D\sb{I} = 128$, $K = 128$, $F = 3$, $P = 1$, and $S = 1$.
Therefore, $W\sb{O} = 32$ and $D\sb{O} = 128$.
In the baseline implementation, each cluster computes one depth slice (aka ``channel'') of the output volume (of dimensions $W\sb{O} \times W\sb{O}$) at a time.
As the entire input volume does not fit into the local memory of a cluster, the cluster loads a stack of depth slices of the input volume at a time.
Thus, each cluster needs to load the entire input volume once per output depth slice.
As the first result column in \cref{tbl:sys:manticore:perf} shows, this implies a very low operational intensity %
and entails that performance is bound by the HBM memory bandwidth. %
One strategy to alleviate this is to let each cluster compute a stack of depth slices of the output volume.
As the input depth slices can be reused for multiple output depth slices, this reduces the amount of data transferred per computation.
For a stack of 8 output depth slices (second column),
the operational intensity is sufficiently high that the performance becomes compute-bound.
To save even more off-chip bandwidth without sacrificing performance, the hierarchical network can be used to form a processing pipeline where clusters obtain 
their input depth slice from another cluster instead of off-chip memory.
The third column shows that when all 16 clusters within one L2 quadrant form such a pipeline, the off-chip memory traffic can be massively reduced while performance is maintained.
Traffic is also reduced on the L2 and L3 networks
because data, once it is in the local memory of a cluster, is mainly transferred through the L1 networks.

A fully-connected \gls{nn} layer transforms an input volume with dimensions $W\sb{I} \times W\sb{I} \times D\sb{I}$ to an output volume with dimensions $1 \times 1 \times D\sb{O}$.
As any fully-connected layer can be represented as convolutional layer, we stick to the introduced notation and operations and set $W\sb{I} = 32$, $D\sb{I} = 128$, $K = 128$, $F = 32$, $P = 0$, and $S = 1$.
Therefore, $W\sb{O} = 1$ and $D\sb{O} = 128$.
As each filter parameter is used only once per input-output volume pair, fully-connected layers are usually implemented by transforming a \emph{batch} of $B$ input volumes to a batch of $B$ output volumes; we use $B = 32$.
Our implementation parallelizes the input depth slices over the clusters.
Before the parallel section, each cluster allocates a private output volume and initializes it to zero.
In the parallel section, the cluster first loads the entire batch of one depth slice of the input volume and then loops over the output depth slices.
Within that loop, the cluster loads the filter parameters for the current pair of input and output depth slices and then enters an inner loop over the batch.
Within the inner loop, the cluster multiplies the input depth slice of a batch element with the loaded filter parameters element-wise and then accumulates all products to a single value, which it adds to the output element for the current output depth slice and batch element.
After the parallel section, the private output volumes of all clusters are reduced by summation to a single output volume, which contains the contributions of all input depth slices.
As the last column in \cref{tbl:sys:manticore:perf} shows, this implementation is compute-bound.
There is no communication between the clusters in the parallel region  because there is no data common to any two clusters.
As such, the hierarchy of the network is not exploited, but the high bandwidth between HBM and any cluster is: it allows reaching compute-boundedness already for a batch size of 32.
Larger batch sizes further reduce the off-chip bandwidth.

\section{Related Work}%
\label{sec:related_work}

\Gls{noc} topologies, routing algorithms, flow control schemes, and router architectures have been subject to a vast amount of research (see \cite{pasricha2010,flich2010,jerger2017,kundu2018nocs,dally2003nocs,benini2006nocs} for detailed reviews).
Important conclusions from this research are that the optimal on-chip network topology highly depends on the target application and computer architecture, and that routing strategies and flow control schemes are intertwined with the communication protocol, which all connected modules need to adhere to.
Thus, we do not try to innovate in this field.
Rather, the modules in our platform allow to build an on-chip network with arbitrary topology that adheres to a state-of-the-art, industry-standard protocol,
following the paradigm put forward by application-specific \gls{noc} research efforts (see \cite{cilardo2016} for an up-to-date survey).
Additionally, our elementary modules allow to design custom network modules\revhl{, including custom endpoints such as caches and memory controllers, without having to deal with all protocol intricacies}.
\revhl{To the best of our knowledge, our work is the first on-chip communication platform that offers elementary modules smaller than crossbars or switches.}

%

Non-coherent on-chip communication is central for heterogeneous, accelerator-rich \glspl{soc}~\cite{giri2018}.
Protocols similar to AMBA \acrshort{axi}5~\cite{axi}, which our platform directly supports, are
IBM's CoreConnect~\cite{coreconnect}, Silicore's Wishbone~\cite{wishbone}, Accellera's Open Core Protocol (OCP)~\cite{accellera:ocp}, and SiFive's TileLink Uncached Heavyweight (TL-UH)~\cite{tilelink}.
They all, like \acrshort{axi}, are royalty-free standards.
CoreConnect, Wishbone, and OCP provide a subset of the features of \acrshort{axi}5, %
and while they had been used in the past, they are nowadays not nearly as widely used as \acrshort{axi}.
TL-UH, like \acrshort{axi}5, supports burst transactions, multiple outstanding transactions, and transaction reordering and uses valid-ready flow control.
TL-UH has stricter forward progress requirements than \acrshort{axi}5, which our modules could also fulfill.
While the specifications define protocols for on-chip communication, they do not describe the architecture of network modules implementing them; that is an important contribution of our work.
The OpenSoC~Fabric~\cite{fatollahi2016} is an open-source implementation of a custom non-coherent protocol, with an interface to AXI-Lite in development.
AXI-Lite does not support bursts or transaction reordering and is therefore not suited for high-performance communication.
The ESP project~\cite{giri2018nocs} provides an open-source implementation of a 2D-mesh \gls{noc} with a custom protocol.
\revhl{%
  Similarly, the non-coherent BaseJump Manycore Accelerator Network, which has first been used in the Celerity chip~\cite{davidson2018}, adheres to a custom protocol and is designed for 2D-mesh networks.
}
In contrast, our platform is topology-agnostic and adheres to an industry-standard protocol.
An overview of on- and off-chip interconnects for \gls{nn} accelerators is presented in \cite{nabavinejad2020DNNnetworks}.
They highlight the need for non-mesh topologies in \gls{nn} accelerators, to which we contribute with our case study and topology-independent platform.

\colorlet{colorgood}{OliveGreen}
\colorlet{colorokay}{Orange}
\colorlet{colorquitegood}{colorgood!40!colorokay}
\colorlet{colorbad}{OrangeRed}
\colorlet{colorunknown}{Gray!60!black}

\newcommand{\good}[1]{\textcolor{colorgood}{#1}}
\newcommand{\quitegood}[1]{\textcolor{colorquitegood}{#1}}
\newcommand{\okay}[1]{\textcolor{colorokay}{#1}}
\newcommand{\bad}[1]{\textcolor{colorbad}{#1}}

\newcommand\circledsym[2]{%
  \adjustbox{height=1.15em,margin*=0 -1.25 -2.5 0}{%
    \tikz\node[circle,color=white,fill=#1,inner sep=.2pt,font=\bfseries]{#2};%
  }
}
\newcommand{\goodyes}{\circledsym{colorgood}{$\pmb\checkmark$}}
\newcommand{\quitegoodyes}{\circledsym{colorquitegood}{$\pmb\checkmark$}}
\newcommand{\badno}{\circledsym{colorbad}{\textsf{X}}}
\newcommand{\unknown}{\circledsym{colorunknown}{?}}
\newcommand{\soldseparately}{\circledsym{BurntOrange}{\textbf{\$}}}

\newcolumntype{R}{%
    >{\adjustbox{right=2.4cm,angle=320,lap=-\width+1.5em}\bgroup}%
    l%
    <{\egroup}%
}
\newcommand*\rot{\multicolumn{1}{R}}%
\newcommand{\tblrottitle}[1]{\rot{\textbf{#1}}}
\newcommand{\productentry}[3]{\parbox[c][2.5em]{6em}{#1\,\tiny\cite{#3}\\\scriptsize#2}}

\begin{table}
  \ifthesis\includegraphics[width=\textwidth]{related_work_table.pdf}\else
  \ifrevhl\color{colorrevhl}\fi%
  \scriptsize%
  \setlength{\tabcolsep}{1pt}%
  \sisetup{range-phrase=--}%
  \resizebox{\columnwidth}{!}{%
  \begin{tabular}{ l c c c c c c c c c c c c c c c }
    \toprule
    \parbox[t][6em][b]{3em}{\textbf{Product}} 
      & \tblrottitle{Public Prod.\ Spec.}
      & \tblrottitle{Open Architecture}
      & \tblrottitle{Disclosable AT Results}
      & \tblrottitle{RTL Open-Source}
      & \tblrottitle{RTL Modifiable}
      & \tblrottitle{Usable on FPGA}
      & \tblrottitle{Usable on ASIC}
      & \tblrottitle{Elems.\ for Custom IPs}
      & \tblrottitle{Data Width [bit]}
      & \tblrottitle{Max.\ Concur.\ Txns.}
      & \tblrottitle{Data Width Conv.}
      & \tblrottitle{ID Width Conv.}
      & \tblrottitle{DMA Engine}
      & \tblrottitle{OCM Controller}
      \\
    \midrule
    \productentry{Arm}{NIC-400}{nic400}
      & \goodyes{}
      & \badno{}
      & \badno{}
      & \badno{}
      & \unknown{}
      & \goodyes{}
      & \goodyes{}
      & \badno{}
      & \bad{\numrange{32}{256}}
      & \good{\numrange{1}{127}}
      & \goodyes{}
      & \quitegoodyes{}\textsuperscript{\dag}
      & \soldseparately{}
      & \soldseparately{}
      \\
    \productentry{Arteris}{FlexNoc}{flexnoc}
      & \badno{}
      & \badno{}
      & \badno{}
      & \badno{}
      & \unknown{}
      & \unknown{}
      & \goodyes{}
      & \badno{}
      & \unknown{}
      & \unknown{}
      & \unknown{}
      & \unknown{}
      & \unknown{}
      & \unknown{}
      \\
    \productentry{Synopsys}{DW\_axi}{synopsys:dw:axi}
      & \badno{}
      & \badno{}
      & \badno{}
      & \badno{}
      & \soldseparately{}
      & \goodyes{}
      & \goodyes{}
      & \badno{}
      & \okay{$\leq$ 512}
      & \unknown{}
      & \unknown{}
      & \unknown{}
      & \soldseparately{}
      & \unknown{}
      \\
    \productentry{Xilinx}{AXI\,LogiCore}{xilinx:axi}
      & \goodyes{}
      & \badno{}
      & \goodyes{}
      & \badno{}
      & \unknown{}
      & \goodyes{}
      & \badno{}
      & \badno{}
      & \quitegood{\numrange{32}{1024}}
      & \okay{\numrange{1}{32}}
      & \goodyes{}
      & \badno{}
      & \goodyes{}
      & \goodyes{}
      \\
    This Work
      & \goodyes{}
      & \goodyes{}
      & \goodyes{}
      & \goodyes{}
      & \goodyes{}
      & \goodyes{}
      & \goodyes{}
      & \goodyes{}
      & \good{\numrange{8}{1024}}*
      & \good{\numrange{1}{128}}\textsuperscript{\P}
      & \goodyes{}
      & \goodyes{}
      & \goodyes{}
      & \goodyes{}
      \\
    \bottomrule
  \end{tabular}}\\[.25em]
  \scriptsize
  \raisebox{-.2ex}{\unknown{}}~=~not disclosed publicly.
  \raisebox{-.2ex}{\soldseparately{}}~=~licensed separately.
  \dag\,Auto-inferred inside crossbar, cannot be customized or instantiated standalone.
  *\,Limited by AXI standard, larger data widths theoretically possible.
  \P\,Range evaluated in this work, larger values theoretically possible.
  \fi
  \caption{Commercial \acrshort{ip} offerings for \acrshort{axi} compared with this work.}%
  \label{tab:related_work:competition}
\end{table}

\revhl{%
Commercial \gls{ip} offerings for \gls{axi} exist from multiple vendors, and we compare with them in \cref{tab:related_work:competition}.
Our work is the only one (1) whose architecture is fully disclosed in literature, (2) whose \gls{rtl} code is open-source and modifiable, enabling, e.g., the exploration of arbitration algorithms with certain guarantees inside standard-compliant networks, and (3) whose \gls{at} characteristics may be disclosed not only on \glspl{fpga}.
Despite its research origin, the implementation behind this work powers on-chip communication of an increasing number of \glspl{asic} (e.g., \cite{manticore,zaruba2019}).
From a technical perspective, our work is the only one that offers elementary modules (i.e., \emph{network} (de)multiplexers) that can be used to build custom AXI-compliant \gls{ip} modules without having to deal with all intricacies of the protocol, instead of only crossbars as finest-granularity modules.
Additionally, our work supports all standard-defined data widths, supports the highest number of concurrent transactions, and comes with communication modules such as ID width converters, a \gls{dma} engine,
\ifllc\else%
  and
\fi
on-chip memory controllers\ifllc\else\footnote{%
  \revhl{Our platform additionally includes a last-level cache (LLC), which is not described in this paper due to space constraints but is available in our open-source repository.}
}\fi,
which are licensed separately or not available at all from commercial vendors.
}

Cache-coherent on-chip communication protocols currently in use include Intel's UltraPath Interconnect~\cite{intelUPI}, AMD's scalable data fabric~\cite{burd2019}, IBM's Power9 on-chip interconnect~\cite{sadasivam2017}, AMBA \gls{ace}~\cite{axi}, AMBA5 \gls{chi}~\cite{chi}, and TileLink Cached (TL-C)~\cite{tilelink}.
\Gls{ace} and TL-C are extensions of \gls{axi} and TL-UH, respectively.
As such, our platform could be extended for coherent communication by adding channels, transactions, and properties defined by these specifications.
\revhl{%
  Under the ordering rules \orderrule{1--3}, coherence transactions cannot use the existing channels of regular transactions:
  In coherent communication, a regular transaction can entail coherence transactions, which must complete before the regular transaction can complete.
  Guaranteeing this would require different ordering rules.
}
The other protocols are standalone specifications with very different properties.
For instance, we refer to \cite{cavalcante2020} for an open-source bridge for connecting to \gls{chi} from \gls{axi}.
With such a bridge, our platform can connect to a coherent system interconnect if needed, possibly extending to multiple chips.
Coherency in on-chip networks has been studied extensively in research, e.g.,~\cite{eisley2006,jerger2008,agarwal2009}. %
A prominent system example is SCORPIO~\cite{daya2014}, where a coherent mesh \gls{noc} interconnects 36 homogeneous cores on a die.
Their work focuses on the \gls{noc} and router architecture for a coherent homogeneous multi-core, while we design an end-to-end non-coherent on-chip communication platform suitable for heterogeneous many-cores.
Generators for cache-coherent on-chip networks have been presented in multiple works:
Open2C~\cite{open2c} contains a library of modules for coherent networks written in Chisel.
The Rocket chip generator~\cite{rocketchip} constructs \glspl{soc} written in Chisel, and the coherent \gls{noc} adheres to TL-C.
OpenPiton~\cite{openpiton} generates tile-based manycore processors with a 2D mesh coherent \gls{noc}.
One tile has an area of \SI{1.17}{\square\milli\meter} %
when targeting IBM's \SI{32}{\nano\meter} \glsknown{soi} process at \SI{1}{\giga\hertz}.
Of the tile area,
%
\SI{2.7}{\percent}
\ifllc\else%
  are occupied
\fi%
by the $5\times5$ \gls{noc} router.
Accounting for one full technology node difference, the equivalent area in \gls{gf22} would be ca.\
%
\SI{80}{\kilo\GE}%
\ifllc
\else%
  , which
\fi%
\ifthesis
(without any virtual channels)
\fi
is about the same size as a $5\times5$ configuration of our crosspoint%
\ifthesis%
 (with up to 16 reorderable IDs)%
\fi%
.
Open2C and OpenPiton implement a custom protocol, which complicates connectivity with third-party modules, whereas we adhere to an industry-dominant protocol.
The modules in our work are implemented in synthesizable SystemVerilog, so they could be integrated into a higher-level generator as well.

\section{Conclusion}%
\label{sec:conclusion}

This first fully open-source\footnote{%
  SystemVerilog source code available under a permissive open-source license at \url{https://github.com/pulp-platform/axi}.
} platform for high-performance on-chip communication enables the construction of heterogeneous many-core and accelerator-rich \glspl{soc} independent of proprietary on-chip networks \glspl{ip}.
The platform advances the technical state of the art through two main contributions:
First, network (de)multiplexers as elementary components make the design and verification of custom network modules substantially easier.
Second, an end-to-end palette of modules from a \gls{dma} engine to on-chip memory controllers, including data and ID width converters, as well as the widest range of data widths and concurrent transactions enables new designs.
For example, we designed and implemented %
a state-of-the art 1024-core \gls{mlt} accelerator in a modern \SI{22}{\nano\meter} technology, where
our communication fabric provides \SI{32}{\tera
\byte\per\second} cross-sectional bandwidth at only \SI{24}{\nano\second} round-trip latency between any two cores.
Future work enabled by our platform includes design space exploration and optimization of on-chip networks, networks designed for stringent application constraints (e.g., arbitration guarantees for real-time execution), and co-integrated cache-coherent and non-coherent networks.

\bibliographystyle{IEEEtran}
\bibliography{IEEEabrv,main}

\newenvironment{biography}[2][example-image-a]{%
  \begin{IEEEbiography}[{%
    \includegraphics[height=.83in,width=.6457in]{#1}%
  }]{#2}%
}{%
  \end{IEEEbiography}%
}
\newcommand{\lucaphd}{He is currently pursuing a PhD degree in the Digital Circuits and Systems group of Prof.\ Benini.}
\newcommand{\ethgrad}[2]{received his BSc and MSc degree in electrical engineering and information technology from ETH Zurich in #1 and #2, respectively.}
\newcommand{\researchinterests}[1]{His research interests include #1.}

\begin{biography}[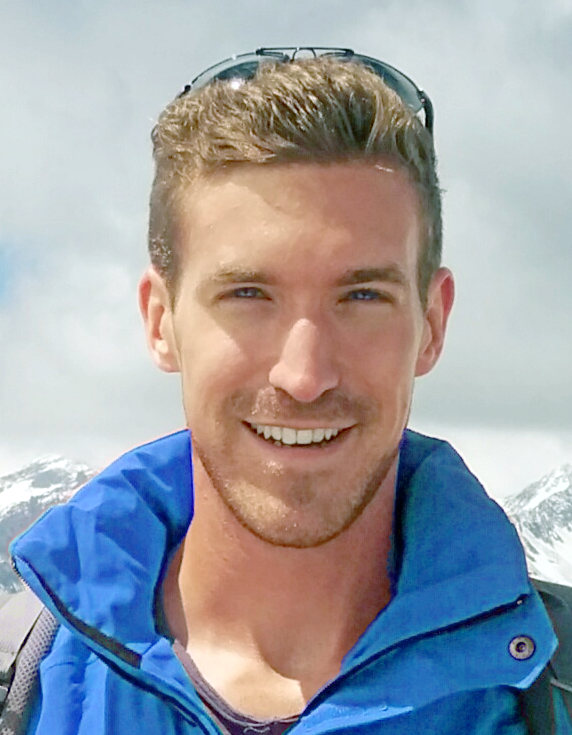]{Andreas Kurth}
    \ethgrad{2014}{2017}
    \lucaphd{}
    \researchinterests{%
        the architecture and programming of heterogeneous \acrshortpl{soc} and %
        accelerator-rich computing systems}
\end{biography}

\begin{biography}[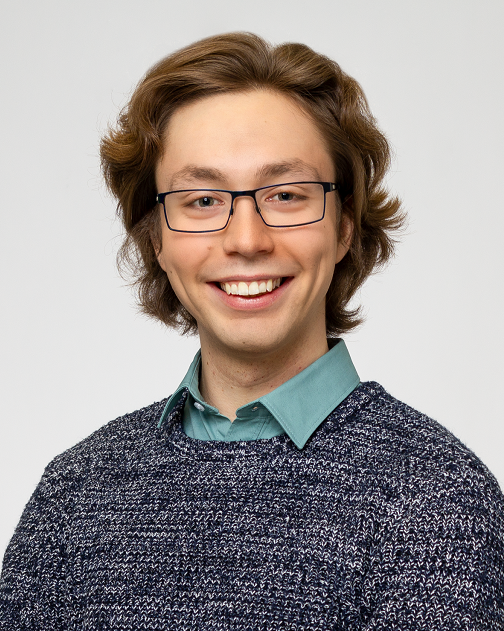]{Wolfgang R\"{o}nninger}
    \ethgrad{2017}{2019}
    He currently works as a research assistant in the Digital Circuits and Systems group of Prof.\ Benini.
    \researchinterests{high-performance on-chip communication networks and general-purpose memory hierarchies}
\end{biography}

\begin{biography}[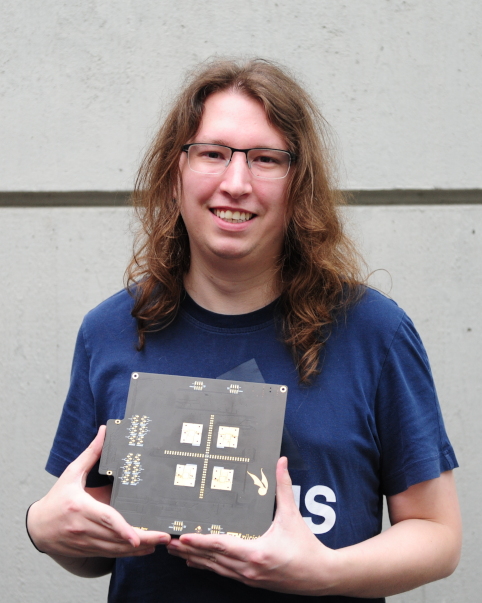]{Thomas Benz}
    \ethgrad{2018}{2020}
    \lucaphd{}
    \researchinterests{energy-efficient high-performance computer architectures and the design of \acrshortpl{asic}}
\end{biography}

\begin{biography}[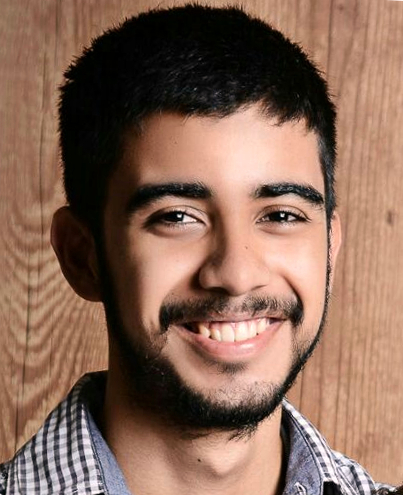]{Matheus Cavalcante}
    received his MSc degree in integrated electronic systems from the Grenoble Institute of Technology (Phelma) in 2018.
    \lucaphd{}
    \researchinterests{vector processing and high-performance computer architectures}
\end{biography}

\begin{biography}[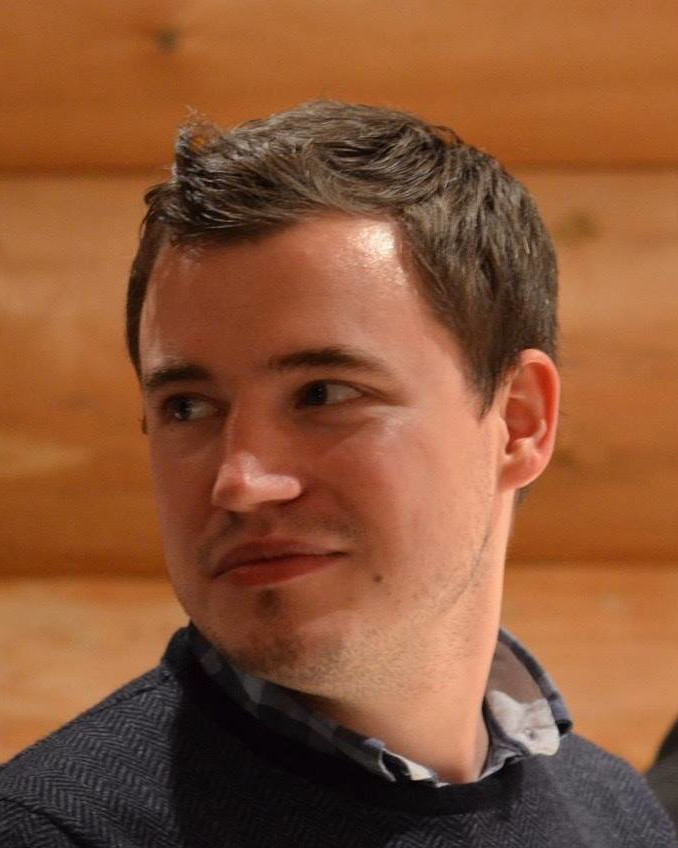]{Fabian Schuiki}
    \ethgrad{2014}{2017}
    \lucaphd{}
    \researchinterests{computer architecture, transprecision computing, as well as near-memory and in-memory processing}
\end{biography}

\begin{biography}[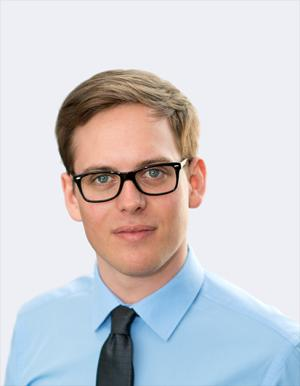]{Florian Zaruba}
    received his BSc degree from TU Wien in 2014 and his MSc from the ETH Zurich in 2017.
    \lucaphd{}
    \researchinterests{design of VLSI circuits and high-performance computer architectures}
\end{biography}

\begin{biography}[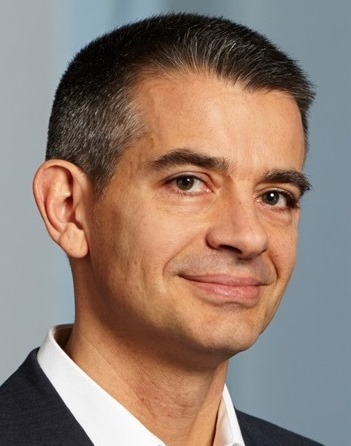]{Luca Benini}
    (F'07) holds the chair of Digital Circuits and Systems at ETH Zurich and is Full Professor at the Università di Bologna.
    Dr.\ Benini’s research interests are in energy-efficient computing systems design, from embedded to high-performance.
    He has published more than 1000 peer-reviewed papers and five books.
    He is a Fellow of the ACM and a member of Academia Europaea.
\end{biography}

\end{document}